\documentclass[prd,twocolumn,showpacs,floatfix,amsmath,nofootinbib,amssymb,floatfix]{revtex4}
\usepackage{graphicx,color,dcolumn,booktabs,bm}
\usepackage{longtable,lscape}
\usepackage{txfonts}
\usepackage{overpic}
\usepackage{amssymb}
\usepackage{indentfirst}
\usepackage{feynmf}   
\usepackage{slashed}  
\usepackage{cases}
\usepackage{color}
\usepackage{multirow}
\usepackage{epstopdf}
\usepackage{longtable}
\usepackage{float}
\usepackage{graphicx,color,dcolumn,booktabs,bm}
\usepackage[colorlinks,
            citecolor=blue,
            anchorcolor=red,
            menucolor=red,
            linkcolor=red,
            filecolor=red,
            runcolor=red,
            urlcolor=blue,
            frenchlinks=red]{hyperref}

\graphicspath{{Figures}} %

\allowdisplaybreaks

\begin{document}

\title{Pion radiative decays of excited hidden-charm pentaquark molecules: from $\Sigma_c^{(*)}\bar{D}^{(*)}(2S)$ molecules to the reported $P_c$ states}

\author{Yu-Jie Tang$^{1}$}
\author{Wen-Yan Peng$^{1,2}$}\email{11636017@zju.edu.cn}
\author{Rui Chen$^{1,2}$}\email{chenrui@hunnu.edu.cn}
\author{Fu-Lai Wang$^{3}$}
\affiliation{
$^1$Key Laboratory of Low-Dimensional Quantum Structures and Quantum Control of Ministry of Education, Department of Physics and Synergetic Innovation Center for Quantum Effects and Applications, Hunan Normal University, Changsha 410081, China\\
$^2$Hunan Research Center of the Basic Discipline for Quantum Effects and Quantum Technologies, Hunan Normal University, Changsha 410081, China\\
$^3$School of Physical Science and Technology, Lanzhou University, Lanzhou 730000, China}
\date{\today}

\begin{abstract}
The discovery of the hidden-charm pentaquarks \(P_c(4312)\), \(P_c(4440)\) and \(P_c(4457)\) by the LHCb Collaboration are very likely to identify as the \(\Sigma_c^{(*)}\bar{D}^{(*)}\) molecules. A natural and crucial extension is the existence of excited molecular partners built from a ground-state charmed baryon and a radially excited anti-charmed meson, namely \(\Sigma_c^{(*)}\bar{D}^{(*)}(2S)\) molecules. In a framework of chiral quark model, we systematic study pion-emission decays of such excited molecules into the known ground-state \(P_c\) molecules. Our results show that the decay widths are sensitive to the spin structures and the coupled-channel interferences, i.e., the \(\Sigma_c\bar{D}(2S)/\Sigma_c\bar{D}^*(2S)/\Sigma_c^*\bar{D}^*(2S)[1/2(1/2^-)]\) state decays to \(P_c(4440)\) with a width of several MeV, while the width to \(P_c(4457)\) is suppressed below \(0.3\) MeV due to destructive interference. The pion-emission decay can be the key to unveiling the excited molecular spectrum of hidden-charm pentaquarks and provides decisive experimental signatures. We expect the future experiments such as the LHCb and PANDA can verify our predictions.

\end{abstract}


\maketitle

\section{introduction}

Exotic hadron states represent a frontier topic in hadron physics. Investigating exotic states cannot only allows us to test and develop non-perturbative Quantum Chromodynamics (QCD) methods (such as lattice QCD, effective field theories, QCD sum rules, and so on) but also provides crucial experimental and theoretical evidence for understanding the essence of the strong interaction and for perfecting the system of hadron spectroscopy.

As a representative of typical exotic hadron states, hidden-charm pentaquarks have seen groundbreaking progress over the past decade. In 2015, the LHCb Collaboration first reported the existence of hidden-charm pentaquark states in the decay process $\Lambda_b^0 \to J/\psi p K^-$, analyzing data from Run 1 of the Large Hadron Collider \cite{LHCb:2015yax}. By examining the invariant mass spectrum of $J/\psi p$, two resonant structures were observed, named $P_c(4380)$ and $P_c(4450)$, respectively. This discovery opened a new chapter in the study of exotic hadrons. Initially, the theoretical community proposed various interpretations for the nature of these two $P_c$ states, including compact pentaquarks, meson-baryon molecular states, and kinematic effects (see the review papers \cite{Chen:2015moa,Karliner:2015ina,Maiani:2015vwa,Lebed:2015tna,Chen:2016qju,Wang:2015ava,Olsen:2017bmm,Dong:2017gaw,Guo:2017jvc,Liu:2019zoy,Brambilla:2019esw,Meng:2022ozq,Wang:2025dur,Bai:2026atm} for further details). Based on the feature that their masses are very close to the thresholds of charmed baryons and anti-charmed mesons, the molecular state interpretation quickly became one of the prevailing views. For instance, in previous work \cite{Chen:2015loa,Chen:2016heh}, we investigated the $\Sigma_c\bar{D}^*$ interaction using the one-boson-exchange (OBE) model and proposed that $P_c(4380)$ and $P_c(4450)$ could be interpreted as loosely bound molecular states composed of a charmed baryon and an anti-charmed meson, respectively.

In 2019, the LHCb Collaboration re-analyzed the full Run 1 and Run 2 data and revealed fine structures that were previously unresolved \cite{LHCb:2019kea}. The results showed that $P_c(4450)$ consists of two overlapping resonances with similar masses, named $P_c(4440)$ and $P_c(4457)$. Simultaneously, a new narrow resonance, $P_c(4312)$, was discovered. The discovery of these three $P_c$ states significantly advanced theoretical interpretations. The most prominent feature is that their masses lie just below the corresponding charmed baryon and anti-charmed meson thresholds: $P_c(4312)$ is close to the $\Sigma_c\bar{D}$ threshold, while $P_c(4440)$ and $P_c(4457)$ are close to the $\Sigma_c\bar{D}^*$ threshold. This feature strongly suggests that they might not be conventional compact pentaquarks but rather hadronic molecules composed of a charmed baryon and an anti-charmed meson \cite{Wu:2010jy,Yang:2011wz,Wang:2011rga,Wu:2012md,Shen:2019evi,Guo:2019fdo,Xiao:2019mvs,He:2019ify,Uchino:2015uha,Chen:2019bip,Yamaguchi:2019seo,Burns:2019iih,Meng:2019ilv,PavonValderrama:2019nbk,Du:2019pij,Wang:2019ato,Xu:2025mhc,Chen:2019asm}. Following the discovery of the three $P_c$ states, our calculations based on the OBE effective potentials and the coupled-channel analysis explicitly indicated that these three states can be consistently interpreted as follows: $P_c(4312)$ corresponds to a $\Sigma_c\bar{D}$ molecule with $I(J^P)=1/2(1/2^-)$, and $P_c(4440)$ and $P_c(4457)$ correspond to $\Sigma_c\bar{D}^*$ molecules with $I(J^P)=1/2(1/2^-)$ and $1/2(3/2^-)$, respectively \cite{Chen:2019asm}.

With the observations of the three $P_c$ structures, a natural question arises: Do excited states of these $P_c$ states exist? If the three reported $P_c$ states are indeed molecules composed of a charmed baryon and a ground-state anti-charmed meson ($\bar{D}$, $\bar{D}^*$), then, following the general systematics of hadron spectra, such molecular systems should possess radial or orbital excitations — similar to the energy level structure of the hydrogen atom from the ground state to excited states in atomic physics. Therefore, it is possible that hidden-charm pentaquark molecular states composed of a ground-state charmed baryon and radially excited anti-charmed mesons exist. For example, in our previous work \cite{Wang:2019nwt}, we adopted the OBE effective potentials to study the interactions between a charmed baryon and an anti-charmed meson in $P$-wave, and predicted the existence of possible excited partners of hidden-charm molecular pentaquarks.

In addition, a more crucial question is: If these excited $P_c$ molecular states exist, how would they decay into the known ground-state $P_c$ states? If molecular states composed of a charmed baryon and an excited anti-charmed meson exist, one of important decay channels would be transitioning to the $P_c$ states composed of a charmed baryon and a ground-state anti-charmed meson by emitting a light meson (particularly a $\pi$ meson) \cite{Xiao:2020frg,Ling:2021lmq}. The physical picture of this process is clear: the excited anti-charmed meson de-excites to a ground-state anti-charmed meson by radiating a $\pi$ meson, while the entire molecular system remains bound, thus completing the transition from an excited molecular state to a ground-state molecular state. This pion radiative decay serves as the most direct probe for studying excited molecular states, one can predict which excited molecular states might be observable experimentally and their dynamical connections to the known $P_c$ states. In our previous work \cite{Sheng:2025rhx}, we calculated the properties within a molecular scenario using the chiral quark model and coupled-channel effects. Our results revealed a strong dependence of the decay widths on the internal structure and spatial wave functions.

In this work, we will study the pion-emission properties of possible $\Sigma_c^{(*)}\bar{D}^{(*)}(2S)$ molecules decaying into the $P_c$ states as the $\Sigma_c^{(*)}\bar{D}^{(*)}$ molecular states. Before calculating the decay widths, we first adopt the OBE effective potentials to predict the mass spectrum of possible $\Sigma_c^{(*)}\bar{D}^{(*)}(2S)$ molecular candidates and obtain the corresponding bound state solutions. In particular, in the heavy quark limit, the OBE effective potentials between the ground-state charmed baryon and the radially excited anti-charmed meson have the same forms as those between the ground-state charmed baryon and the ground-state anti-charmed meson, and because the former has a larger reduced mass, the corresponding binding energy can be a little deeper.

\begin{figure}[!htbp]
    \centering
    \includegraphics[width=0.8\linewidth]{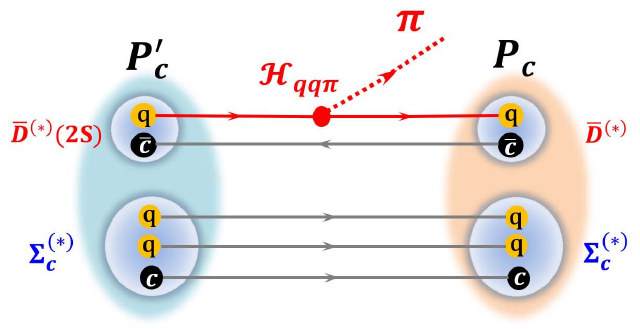}
    \caption{The $\pi$-emission process between the $P_c$ molecular states at tree level.}
    \label{diagram}
\end{figure}

At tree level, the pion-emission process between the excited and ground $P_c$ states is realized through the interaction of their anti-charmed mesons with a pion, as shown in Figure \ref{diagram}. In Refs.~\cite{Xiao:2014ura,Ni:2021pce}, the authors studied the strong decays of radially excited charmed mesons using the chiral quark model, expanding the effective interaction Hamiltonian to the order of the inverse square of the interacting quark masses. The corresponding calculation results are consistent with the experimentally reported decay widths and branching ratios of radially excited charmed mesons \cite{BaBar:2010zpy,LHCb:2013jjb,LHCb:2015eqv,LHCb:2019juy,Wang:2025dur1}. For the above reasons, this work follows the research approach of Refs.~\cite{Xiao:2014ura,Ni:2021pce} to further investigate the pion-emission decay between the excited and ground $P_c$ states by using the chiral quark model \cite{Manohar:1983md,Li:1994cy,Li:1995si,Li:1997gd,Zhao:2002id,Zhong:2007gp,Zhong:2008kd}.

This paper is organized as follows. In Sec.~\ref{sec2}, we derive the pion-emission interactions between $P_c$ states in a molecular scenario. The numerical results are presented and discussed in Sec.~\ref{sec3}. Finally, a summary is provided in Sec.~\ref{sec4}.

\section{Interactions}\label{sec2}

In the chiral quark model, the effective Lagrangians describing the interactions between quarks and pseudoscalar mesons can be constructed as \cite{Manohar:1983md}
\begin{equation}\label{effective coupling}
\mathcal{L}=\frac{\delta}{\sqrt{2}f_{\pi}^{'}}\bar{\psi}\gamma_{\mu}\gamma_5\psi\vec{I}\cdot\partial^\mu\vec{\phi}_{m}.
\end{equation}
Here, $f_{\pi}^{'}=0.093$ GeV and $\delta=0.557$ \cite{Ni:2023lvx,Zhong:2024mnt} are the pseudoscalar meson decay constant and the global parameter accounting for the strength of the quark-pseudoscalar-meson couplings, respectively. $\vec{I}$ is an isospin operator. For the pion-emission interactions \cite{Zhong:2008kd,Zhong:2007gp}, one has $I=a^\dagger (u) a (d)$ for $\pi^-$, $I=a_j^\dagger (d) a_j (u)$ for $\pi^+$, and $I=\dfrac{1}{\sqrt{2}} \left[ a^\dagger (u) a (u) - a^\dagger (d) a (d) \right]$, with $a^\dagger (u,d)$ and $a (u,d)$ being the creation and annihilation operators for $u$ and $d$ quarks, respectively. $\psi$ and $\phi_m$ are the light quark field and the pseudoscalar meson field, respectively.

\renewcommand\tabcolsep{0.35cm}
\renewcommand{\arraystretch}{2.0}
\begin{table*}[!htpb]
\centering
\caption{Flavor wave functions and the spin-orbit wave functions $|{}^{2S+1}L_J\rangle$ for the considered $\Sigma_{c}^{(*)}\bar D$ systems. Here, $I$ and $I_3$ are their isospin and third component, respectively.}\label{function}
{\begin{tabular}{cl|cl}\toprule[1.0pt]\toprule[1.0pt]
$|I,I_3\rangle$    & Configurations     &$J^P$      &{Channels $\left|{}^{2S+1}L_J\right\rangle$}\\\midrule[1.0pt]
$|\frac{3}{2},\frac{3}{2}\rangle$  & $\left|\Sigma_c^{(*)++}{D}^{-}\right\rangle$
   &$1/2^-$ &$\Sigma_c\bar{D}\left|{}^2\mathbb{S}_{\frac{1}{2}}\right\rangle$,\quad $\Sigma_c^*\bar{D}\left|{}^4\mathbb{D}_{\frac{1}{2}}\right\rangle$,\quad $\Sigma_c\bar{D}^*\left|{}^2\mathbb{S}_{\frac{1}{2}}/{}^4\mathbb{D}_{\frac{1}{2}}\right\rangle$,\quad $\Sigma_c^*\bar{D}^*\left|{}^2\mathbb{S}_{\frac{1}{2}}/{}^4\mathbb{D}_{\frac{1}{2}}/{}^6\mathbb{D}_{\frac{1}{2}}\right\rangle$\\
$|\frac{3}{2},\frac{1}{2}\rangle$ & $\sqrt{\frac{1}{3}}\left|\Sigma_c^{(*)++}{D}^{-}\right\rangle+\sqrt{\frac{2}{3}}\left|\Sigma_c^{(*)+}\bar{D}^{0}\right\rangle$
   &$3/2^-$  & $\Sigma_c^*\bar{D}\,\left|{}^4\mathbb{S}_{\frac{3}{2}}/{}^4\mathbb{D}_{\frac{3}{2}}\right\rangle$,\quad $\Sigma_c\bar{D}^*\,\left|{}^4\mathbb{S}_{\frac{3}{2}}/ {}^2\mathbb{D}_{\frac{3}{2}}/{}^4\mathbb{D}_{\frac{3}{2}}\right\rangle$,\quad $\Sigma_c^*\bar{D}^*\,\left|{}^4\mathbb{S}_{\frac{3}{2}}/{}^2\mathbb{D}_{\frac{3}{2}}/ {}^4\mathbb{D}_{\frac{3}{2}}/ {}^6\mathbb{D}_{\frac{3}{2}}\right\rangle$\\
$|\frac{3}{2},-\frac{1}{2}\rangle$ & $\sqrt{\frac{2}{3}}\left|\Sigma_c^{(*)+}{D}^{-}\right\rangle +\sqrt{\frac{1}{3}}\left|\Sigma_c^{(*)0}\bar{D}^{0}\right\rangle$
   &$5/2^-$ &$\Sigma_c^*\bar{D}^*\,\left|{}^6\mathbb{S}_{\frac{5}{2}}/{}^2\mathbb{D}_{\frac{5}{2}}/ {}^4\mathbb{D}_{\frac{5}{2}}/ {}^6\mathbb{D}_{\frac{5}{2}}\right\rangle$\\\cline{3-4}
$|\frac{3}{2},-\frac{3}{2}\rangle$ & $\left|\Sigma_c^{(*)0}{D}^{-}\right\rangle$
    &\multicolumn{2}{l}{$\Sigma_c^*\bar{D}:\,\, \left|{}^{2S+1}L_{J}\right\rangle =
\sum_{m_S,m_L}C^{J,M}_{\frac{3}{2}m_S,Lm_L}
          \Phi_{\frac{3}{2}m_S}|Y_{L,m_L}\rangle$}\\
$|\frac{1}{2},\frac{1}{2}\rangle$ & $\sqrt{\frac{2}{3}}\left|\Sigma_c^{(*)++}{D}^{-}\right\rangle-\sqrt{\frac{1}{3}}\left|\Sigma_c^{(*)+}\bar{D}^{0}\right\rangle$
    &\multicolumn{2}{l}{$\Sigma_c\bar{D}^*: \left|{}^{2S+1}L_{J}\right\rangle =
\sum_{m,m'}^{m_S,m_L}C^{S,m_S}_{\frac{1}{2}m,1m'}C^{J,M}_{Sm_S,Lm_L}
          \chi_{\frac{1}{2}m}\epsilon^{m'}\left|Y_{L,m_L}\right\rangle$}\\
$|\frac{1}{2},-\frac{1}{2}\rangle$ & $\sqrt{\frac{1}{3}}\left|\Sigma_c^{(*)+}{D}^{-}\right\rangle-\sqrt{\frac{2}{3}}\left|\Sigma_c^{(*)0}\bar{D}^{0}\right\rangle$
   &\multicolumn{2}{l}{$\Sigma_c^*\bar{D}^*: \left|{}^{2S+1}L_{J}\right\rangle =
\sum_{m,m'}^{m_S,m_L}C^{S,m_S}_{\frac{3}{2}m,1m'}C^{J,M}_{Sm_S,Lm_L}
          \Phi_{\frac{3}{2}m}\epsilon^{m'}\left|Y_{L,m_L}\right\rangle$}\\
\bottomrule[1.0pt]\bottomrule[1.0pt]
\end{tabular}}
\end{table*}

By expanding the above Lagrangian in a non-relativistic form up to the mass order of $1/m^2$, one can obtain \cite{Arifi:2021orx,Arifi:2022ntc}
\begin{align}\label{form}
\mathcal{H}_{qq\pi} &= g\left(\mathcal{G}\boldsymbol{\sigma}\cdot\boldsymbol{q}
+\frac{\omega_{m}}{2\mu_{q}}\boldsymbol{\sigma}\cdot\boldsymbol{p}\right )F(\boldsymbol{q}^2)I\varphi \nonumber\\
&\quad -\frac{g}{32\mu_{q}^{2}}\left[m_{\mathbb{P}}^2(\boldsymbol{\sigma}\cdot\boldsymbol{q})+2\boldsymbol{\sigma}\cdot(\boldsymbol{q}-2\boldsymbol{p})\times(\boldsymbol{q}\times\boldsymbol{p})\right] \nonumber\\
&\quad \times F(\boldsymbol{q}^2)I\varphi.
\end{align}
Here, we define
\begin{align}
 g &= \delta\sqrt{(E_f+M_f)}/(\sqrt{2}f_{\pi}^{'}),\\
 \mathcal{G} &= -\left(\frac{\omega_{m}}{E_f+M_f}+1+\frac{\omega_{m}}{2m'}\right),\\
 \mu_q &= \frac{mm^{\prime}}{m+m^{\prime}}.
\end{align}
$\boldsymbol{\sigma}$ and $\boldsymbol{p}$ correspond to the spin operator and internal momentum operator of the light quark, respectively. $\varphi=e^{-i\boldsymbol{q}\cdot\boldsymbol{r}}$ is the plane wave part of the emitted light meson. $(E_i, M_i)$ and $(E_f, M_f)$ are the energy and mass of the initial and final particles involved in the interaction, respectively. $\omega_{m}$, $\boldsymbol{q}$, and $m_{\mathbb{P}}$ are the energy, three-momentum, and mass of the emitted pseudoscalar meson. $m$ and $m^{\prime}$ stand for the masses of the initial and final quarks, respectively. In Refs.~\cite{Zhong:2024mnt,Ni:2023lvx}, the quark masses are taken as $m_{u,d}=0.45$ GeV and $m_{c}=1.68$ GeV. In order to suppress unphysical contributions in the high-momentum region, we introduce a form factor $F(\boldsymbol{q}^2)=\sqrt{{\Lambda^2}/{\left(\Lambda^2+\boldsymbol{q}^2\right)}}$ at the interaction vertex, and the cutoff $\Lambda$ is taken to be of the order of the chiral symmetry breaking scale. In Refs.~\cite{Zhong:2024mnt,Ni:2023lvx,Yu:1995ag,Vijande:2004he,Valcarce:2008dr}, the cutoff parameter $\Lambda$ is taken as $\Lambda=0.66$ GeV. Here, we also notice that the different form factors in the OBE potential and the pion-emission vertex reflect their distinct roles: the former regularizes the off-shell exchanged mesons at the hadron level, while the latter suppresses high-momentum components in the quark-level transition operator. Both are standard choices in their respective contexts. We have also tested the stability of our results by using the same monopole form factor for the pion vertex with $\Lambda\sim 1$ GeV; the decay widths change by less than a factor of two, and the interference patterns are unaffected. Therefore, our predictions are stable against the regulator choice.

With the above preparations, we then calculate the pion-emission width between the excited and ground hidden-charm molecular states. In this work, we consider only the pion emission from the anti-charmed meson constituent, which is the dominant mechanism. In the heavy-quark limit, the $\bar{D}^{(*)}\bar{D}^{(*)}\pi$ coupling is leading order, while the $\Sigma_c\Sigma_c\pi$ vertex is suppressed by heavy-quark spin-flip or vanishes in the exact SU(2) and heavy-quark symmetry limits. Contributions from baryon emission or final-state rescattering are higher-order effects and are not expected to alter the order-of-magnitude estimates and the interference patterns reported here. This approximation is consistent with previous works on $P_c$ pionic decays \cite{Xiao:2020frg,Ling:2021lmq,Sheng:2025rhx,Xiao:2014ura,Ni:2021pce}.

In the rest frame of the initial state, the decay width for the $P_c^{\prime}\to P_c+\pi$ process is given by
\begin{align}
\Gamma = \frac{1}{8\pi} \frac{|\boldsymbol{q}|}{M_{P_c^{\prime}}^2} \frac{1}{2J^{P_c^{\prime}} + 1} \sum_{J^{P_c}_z} \left| \mathcal{M}_{J^{P_c^{\prime}}_z J^{P_c}_z}\left(P_c^{\prime}\to P_c+\pi\right) \right|^2,
\label{decay_width}
\end{align}
where $J$ and $J_z$ are the total angular momentum quantum number and its projection, respectively. $\bm{q}=\sqrt{\left(M_{P_c^{\prime}}^2-(M_{P_c}+M_{\pi})^2\right)\left(M_{P_c^{\prime}}^2-(M_{P_c}-M_{\pi})^2\right)}/(2M_{P_c^{\prime}})$ is the three-momentum of the final states, with $M_{A}$ being the mass of particle $A$. $\mathcal{M}\left(P_c^{\prime}\to P_c+\pi\right) = \left\langle \Phi_{P_c} \left| \sum_q\mathcal{H}_{qq\pi} \right| \Phi_{P_c^{\prime}} \right\rangle$ is the decay amplitude for the process $P_c^{\prime}\to P_c+\pi$, where $|\Phi_{P_c^{\prime}}\rangle$ and $|\Phi_{P_c}\rangle$ are the wave functions of the initial and final hidden-charm molecular pentaquarks, respectively.

The wave function of molecular pentaquarks can be expressed as the direct product of the wave functions of their constituent hadrons and the wave function between the constituents $|\Phi\rangle$, i.e., $|\Phi_{P_c}\rangle = |\Phi_M\rangle\otimes|\Phi_{B}\rangle\otimes|\Phi\rangle$. Here, $\Phi_M$ and $\Phi_B$ stand for the wave functions of the constituent meson and baryon, respectively, which include the color wave functions $|\phi_c\rangle=1$, the flavor wave functions $|I,I_3\rangle$, the spin-orbit wave functions $|{}^{2S+1}L_J\rangle$, and the spatial wave functions $|\phi(r)\rangle$. In Table~\ref{function}, we summarize the flavor wave functions $|I,I_3\rangle$ and the spin-orbit wave functions $|{}^{2S+1}L_J\rangle$ for the $P_c$ molecules. When constructing the general expressions of the spin-orbit wave functions for the $\Sigma_c^{(*)}\bar{D}^{(*)}$ system, we define several useful notations, namely $C^{J,M}_{Sm_S,Lm_L}$, $C^{S,m_S}_{\frac{1}{2}m,1m'}$, and $C^{S,m_S}_{\frac{3}{2}m,1m'}$ as the Clebsch-Gordan coefficients. $\chi_{\frac{1}{2}m}$ and $Y_{L,m_L}$ stand for the spin wave function and the spherical harmonic functions, respectively. The polarization vector $\epsilon$ for the $\bar{D}^*$ vector meson is defined as $\epsilon_{\pm}^{m}=\mp\frac{1}{\sqrt{2}}\left(\epsilon_x^{m}{\pm}i\epsilon_y^{m}\right)$ and $\epsilon_0^{m}=\epsilon_z^{m}$, which satisfy $\epsilon_{\pm1}= \frac{1}{\sqrt{2}}\left(0,\pm1,i,0\right)$ and $\epsilon_{0} =\left(0,0,0,-1\right)$. The polarization tensor $\Phi_{\frac{3}{2}m}$ for $\Sigma_c^*$ is expressed as $\Phi_{\frac{3}{2}m}=\sum_{m_1,m_2}\langle\frac{1}{2},m_1;1,m_2|\frac{3}{2},m\rangle\chi_{\frac{1}{2},m_1}\epsilon^{m_2}$.

In this work, we use simple harmonic oscillator (SHO) wave functions to describe the internal spatial wave functions of the mesons and baryons \cite{Li:2008xy,Liu:2011yp,Lai:2024jfe,Zhang:2025ame,Wang:2022nqs}:
\begin{align}
\phi_{n,l,m}(\beta, \mathbf{r}) =& \sqrt{\frac{2n!}{\Gamma\left(n + l + \frac{3}{2}\right)}} L_n^{l + \frac{1}{2}} \left( \beta^2 r^2 \right) \beta^{l + \frac{3}{2}} \mathrm{e}^{-\frac{\beta^2 r^2}{2}} r^l Y_{lm}(\Omega).
\label{SHO}
\end{align}
Here, $L_n^{l + \frac{1}{2}}$ is the associated Laguerre polynomial. Following the convention in Refs.~\cite{Barnes:2002mu,Ackleh:1996yt,Barnes:1996ff}, the parameter $\beta$ is taken as $0.40$ GeV. For the wave functions between the constituents $|\Phi\rangle$, we obtain them by numerically solving the Schr\"{o}dinger equations with the OBE effective potentials. Specifically, for the reported $P_c$ states ($P_c(4312)$, $P_c(4380)$, $P_c(4440)$, and $P_c(4457)$), we reproduce their central masses. For the remaining molecular states, we vary the cutoff values of the monopole-type form factor in the OBE effective potentials to obtain binding energies in the region $E > -15$ MeV. In the following, we present the binding-energy dependence of the pion-emission widths between the excited and ground $P_c$ molecules.

It is worth clarifying the consistency of our hybrid approach. The OBE model describes the hadron-level interactions (including both long-range pion exchange and shorter-range $\rho/\omega$ exchanges) that bind the molecular states, while the chiral quark model provides a microscopic description of the quark-level $\pi$-emission vertex, including the $1/m^2$ relativistic corrections essential for the node structure of radial excitations. Both frameworks are constrained by chiral symmetry, but they constitute different effective theories with different degrees of freedom. This separation of scales is standard in the literature (see, e.g., Refs.~ \cite{Sheng:2025rhx,Wang:2022nqs,Lai:2024jfe}) and is justified because the decay vertex involves short-distance quark dynamics that probes the internal structure of the $2S$ mesons, which the purely hadronic OBE framework cannot easily describe without introducing an unknown $2S \to 1S$ transition form factor. In contrast, the parameters of the chiral quark model (constituent quark masses, oscillator parameter $\beta$, and cutoffs) are fixed from independent observables such as $f_\pi$ and measured hadron radii.

\section{Numerical results}\label{sec3}

\begin{figure*}[!htbp]
\includegraphics[width=1\linewidth]{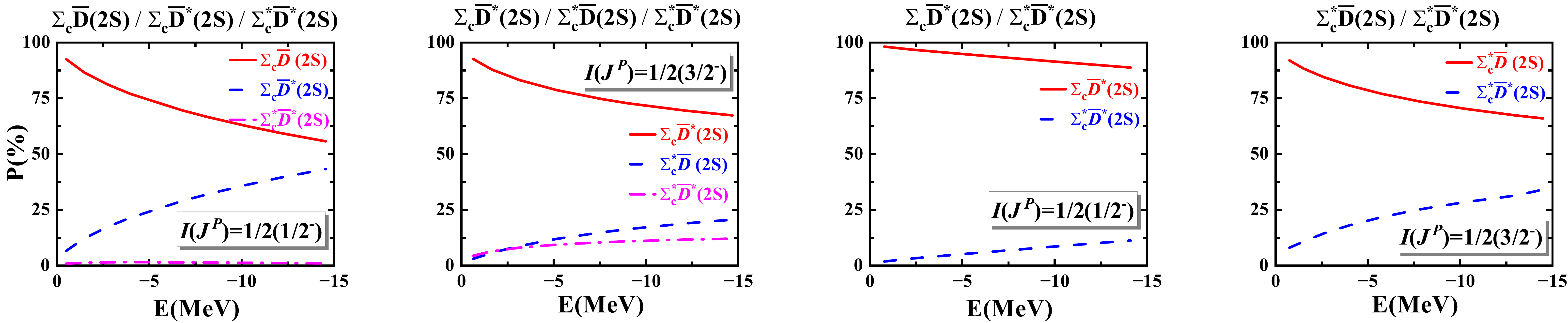}
\caption{The binding energies dependence of the probabilities $P$ (\%) for the involved channels of the isodoublet coupled $\Sigma_c^{(*)}\bar{D}^{(*)}(2S)$ molecules.}
\label{states}
\end{figure*}

By employing the same OBE effective potentials (as shown in Appendix \ref{app01}) as those used for the $\Sigma_c^{(*)}\bar{D}^{(*)}$ systems, we find that molecular candidates for $\Sigma_c^{(*)}\bar{D}^{(*)}(2S)$ indeed emerge when the cutoff parameter in the monopole form factor of the effective potential is taken within a reasonable range, namely $\Lambda \sim 1.00$ GeV. These candidates include (i) the $\Sigma_c\bar{D}(2S)/\Sigma_c\bar{D}^*(2S)/\Sigma_c^*\bar{D}^*(2S)$ molecular state with $I(J^P)=1/2(1/2^-)$; (ii) the $\Sigma_c\bar{D}^*(2S)/\Sigma_c^*\bar{D}(2S)/\Sigma_c^*\bar{D}^*(2S)$ molecular states with $I(J^P)=1/2(1/2^-,3/2^-)$; (iii) the $\Sigma_c^*\bar{D}(2S)/\Sigma_c^*\bar{D}^*(2S)$ molecular state with $I(J^P)=1/2(3/2^-)$; and (iv) the $\Sigma_c^*\bar{D}^{*}(2S)$ molecular states with $I(J^P)=1/2(1/2^-,3/2^-,5/2^-)$. In the Appendix \ref{app01}, we present the corresponding OBE effective potentials together with the dependence of the binding energies and root-mean-square radii on the cutoff parameter. In Figure \ref{states}, we present the fractions of individual channels in these molecular states as functions of the binding energy. We can conclude that the coupled-channel effects play an essential role in the formation of these molecular states. Subsequently, we also introduce the coupled-channel approach to compute the pion-emission decay widths from the excited $P_c$ molecular states to the ground-state $P_c$ molecular states.

\subsection{$\Sigma_c\bar{D}(2S)/\Sigma_c\bar{D}^*(2S)/\Sigma_c^*\bar{D}^*(2S)[{1}/{2}({1}/{2}^-)] \to P_c+ \pi$}

We first compute the decay widths for the following processes:
\begin{itemize}
\item $\Sigma_c\bar{D}(2S)/\Sigma_c\bar{D}^*(2S)/\Sigma_c^*\bar{D}^*(2S)[1/2(1/2^-)]$ decaying into $P_c(4312) + \pi$,
\item $\Sigma_c\bar{D}(2S)/\Sigma_c\bar{D}^*(2S)/\Sigma_c^*\bar{D}^*(2S)[1/2(1/2^-)]$ decaying into $P_c(4380) + \pi$,
\item $\Sigma_c\bar{D}(2S)/\Sigma_c\bar{D}^*(2S)/\Sigma_c^*\bar{D}^*(2S)[1/2(1/2^-)]$ decaying into $P_c(4440) + \pi$,
\item $\Sigma_c\bar{D}(2S)/\Sigma_c\bar{D}^*(2S)/\Sigma_c^*\bar{D}^*(2S)[1/2(1/2^-)]$ decaying into $P_c(4457) + \pi$.
\end{itemize}

\begin{figure}[!htbp]
\includegraphics[width=1\linewidth]{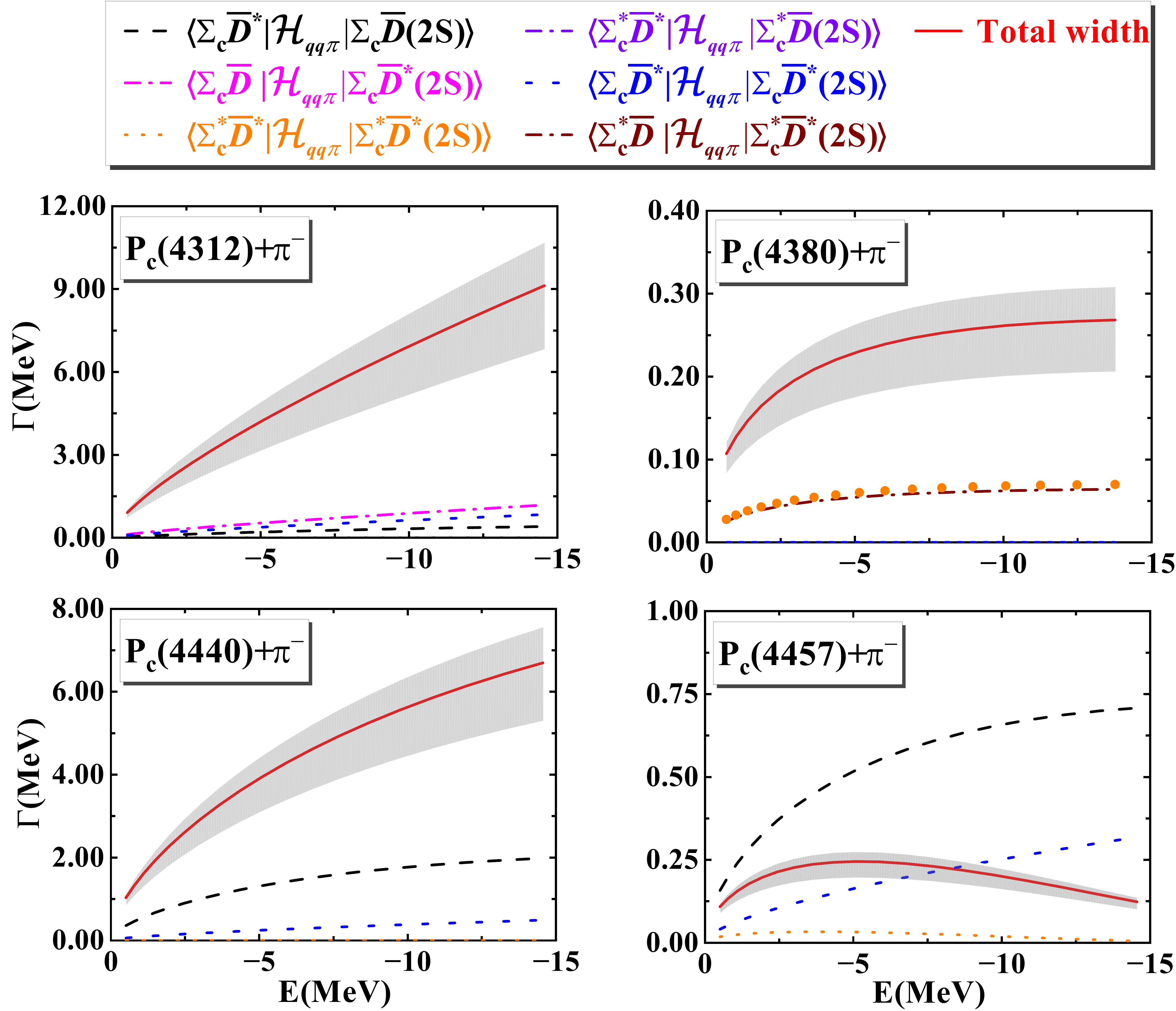}
\caption{The pion-emission widths for $\Sigma_c\bar{D}(2S)/\Sigma_c\bar{D}^*(2S)/\Sigma_c^*\bar{D}^*(2S)$ molecules with $I(J^P)=1/2(1/2^-)$ decaying into $P_c$ molecular states. Here, $E$ is the binding energy for the initial molecular states. $\langle A|\mathcal{H}_{qq\pi}|B\rangle$ corresponds to the pion-emission width for $B$ decaying to $A$. The cutoff values corresponding to the gray shaded area of the data are taken from 0.50 to 0.80 GeV, while the decay widths in the curve manner are calculated at $\Lambda=0.66$ GeV.}
\label{fignum1}
\end{figure}

The corresponding decay amplitude can be expressed as
\[
\begin{aligned}
\mathcal{M} &= \mathcal{M}_1\bigl(\Sigma_c\bar{D}(2S)\to\Sigma_c\bar{D}^*+\pi\bigr) \\
&\quad + \mathcal{M}_2\bigl(\Sigma_c\bar{D}^*(2S)\to\Sigma_c\bar{D}^*+\pi\bigr) \\
&\quad + \mathcal{M}_3\bigl(\Sigma_c\bar{D}^*(2S)\to\Sigma_c\bar{D}+\pi\bigr) \\
&\quad + \mathcal{M}_4\bigl(\Sigma_c^*\bar{D}(2S)\to\Sigma_c^*\bar{D}^*+\pi\bigr) \\
&\quad + \mathcal{M}_5\bigl(\Sigma_c^*\bar{D}^*(2S)\to\Sigma_c^*\bar{D}+\pi\bigr) \\
&\quad + \mathcal{M}_6\bigl(\Sigma_c^*\bar{D}^*(2S)\to\Sigma_c^*\bar{D}^*+\pi\bigr).
\end{aligned}
\]
In Figure \ref{fignum1}, we present the dependence of the pion-emission width on the initial-state binding energy $E$. To assess the theoretical uncertainty, we vary the cutoff $\Lambda$ in the quark-meson form factor of Eq. (\ref{form}) within the range $0.5$–$0.8$ GeV, here, the relevant decay widths are labeled in the gray area. We can see that the absolute decay widths vary by less than 2 times, and the relative hierarchy among different final states and the interference patterns remain unchanged. Moreover, these conclusions remain unchanged in the subsequent research systems.  Thus, our main conclusions are robust against reasonable variations of the phenomenological parameters.

For binding energies $E > -15$~MeV, the width for pion-emission from the $\Sigma_c\bar{D}(2S)/\Sigma_c\bar{D}^*(2S)/\Sigma_c^*\bar{D}^*(2S)[1/2(1/2^-)]$ molecular state to $P_c(4312)$ is on the order of a few MeV, and the decay width increases as the binding energy decreases. As shown in Figure \ref{fignum1}, we also show the decay widths for each individual channel of the initial molecular state decaying to $P_c(4312)$ and a pion. The three processes $\Sigma_c\bar{D}^*(2S)\to\Sigma_c\bar{D}+\pi$, $\Sigma_c\bar{D}^*(2S)\to\Sigma_c\bar{D}^*+\pi$, and $\Sigma_c\bar{D}(2S)\to\Sigma_c\bar{D}^*+\pi$ are almost equally important for the decay $\Sigma_c\bar{D}(2S)/\Sigma_c\bar{D}^*(2S)/\Sigma_c^*\bar{D}^*(2S)[1/2(1/2^-)]\to P_c(4312)+\pi$. The main reason is that $\Sigma_c\bar{D}^*(2S)$ and $\Sigma_c\bar{D}^*$ account for significant proportions in the initial and final molecular states, respectively; in particular, as the binding energy of the initial molecular state increases, the proportion of the $\Sigma_c\bar{D}^*(2S)$ component can reach nearly 50\%. In addition, although the initial and final molecular states are dominated by $\Sigma_c\bar{D}(2S)$ and $\Sigma_c\bar{D}$, respectively, the process $\Sigma_c\bar{D}(2S)\to\Sigma_c\bar{D}+\pi$ is forbidden by spin-parity and thus does not contribute to the total width.

For the $\Sigma_c\bar{D}(2S)/\Sigma_c\bar{D}^*(2S)/\Sigma_c^*\bar{D}^*(2S)[1/2(1/2^-)] $ decaying into $ P_c(4440)/P_c(4457)+\pi$, since the $\Sigma_c\bar{D}(2S)$ and $\Sigma_c\bar{D}^*$ components dominate the initial and final molecular states, respectively, the $\Sigma_c\bar{D}(2S) \to \Sigma_c\bar{D}^* + \pi$ interaction plays an important role in these two decay processes. As shown in Figure \ref{fignum1}, when the binding energy satisfies $E > -15$~MeV, the pion-emission width of the $\Sigma_c\bar{D}(2S)/\Sigma_c\bar{D}^*(2S)/\Sigma_c^*\bar{D}^*(2S)[1/2(1/2^-)]$ molecular state to $P_c(4440)$ can reach several MeV, which is significantly larger than that to $P_c(4457)$ (less than 0.3~MeV). The main reason is that, after incorporating coupled-channel effects, the $\Sigma_c\bar{D}(2S) \to \Sigma_c\bar{D}^* + \pi$ interaction interferes constructively with other contributions for the $P_c(4440)$ final state, whereas it interferes destructively for the $P_c(4457)$ final state.

The decay width of the same initial molecular state to $P_c(4380)$ and a pion is around 0.25~MeV. The small width arises from two factors: first, the dominant $\Sigma_c^*\bar{D}$ component of $P_c(4380)$ cannot be directly accessed from the initial state without a heavy-quark spin flip, and the only contributing channel $\Sigma_c^*\bar{D}^*(2S)\to\Sigma_c^*\bar{D}+\pi$ is suppressed by the very small (a few percent) $\Sigma_c^*\bar{D}^*(2S)$ component in the initial molecular state; second, the remaining coupled-channel contributions involve components that also play only a minor role in the wave functions.

In addition, to assess the theoretical uncertainty, we vary the cutoff $\Lambda$ in the quark-meson form factor of Eq. (\ref{form}) within the range $0.5$–$0.8$ GeV, following the standard values used in the chiral quark model. As shown in Figure \ref{fignum1}, although the absolute decay widths vary by less than 2 times, the relative hierarchy among different final states and the interference patterns remain unchanged. Thus, our main conclusions are robust against reasonable variations of the phenomenological parameters.

\subsection{$\Sigma_c\bar{D}^*(2S)/\Sigma_c^*\bar{D}^*(2S)[{1}/{2}({1}/{2}^-)] \to P_c+ \pi$}

\begin{figure}[!htbp]
\includegraphics[width=1\linewidth]{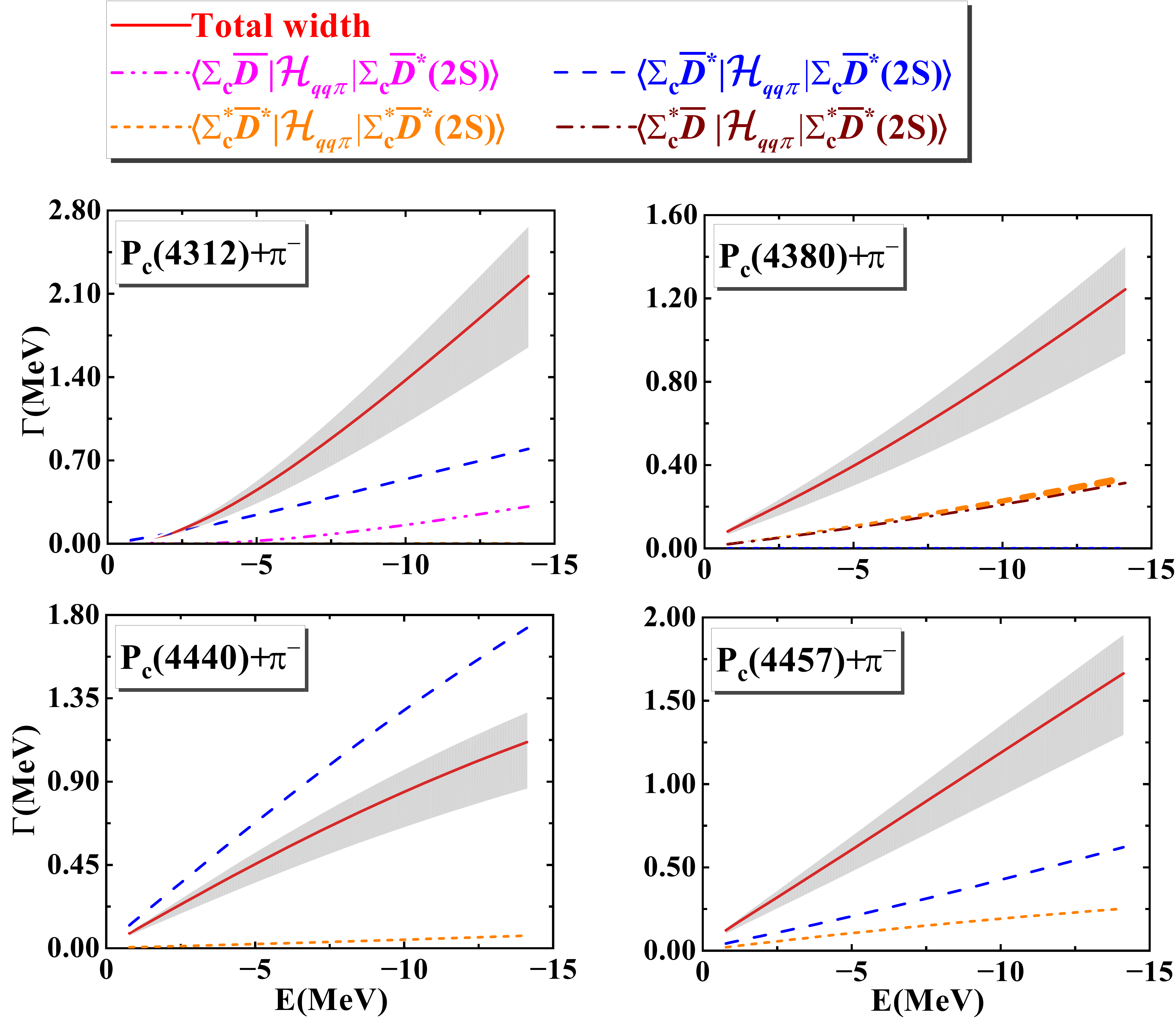}
\caption{The pion-emission widths for $\Sigma_c\bar{D}^*(2S)/\Sigma_c^*\bar{D}^*(2S)$ molecules with $I(J^P)=1/2(1/2^-)$ decaying into $P_c$ molecular states. Here, $E$ is the binding energy for the initial molecular states. $\langle A|\mathcal{H}_{qq\pi}|B\rangle$ corresponds to the pion-emission width for $B$ decaying to $A$. The cutoff values corresponding to the gray shaded area of the data are taken from 0.50 to 0.80 GeV, while the decay widths in the curve manner are calculated at $\Lambda=0.66$ GeV.}\label{fignum2}
\end{figure}

Next, we calculate the widths for the pion-emission transitions from the $\Sigma_c\bar{D}^*(2S)/\Sigma_c^*\bar{D}^*(2S)[1/2(1/2^-)]$ states to $P_c(4312)$, $P_c(4380)$, $P_c(4440)$, and $P_c(4457)$. In Figure \ref{fignum2}, we show the dependence of the pion-emission width on the binding energy of the initial molecular state. We find that when the binding energy $E > -15$~MeV, the pion-emission decay widths of the $\Sigma_c\bar{D}^*(2S)/\Sigma_c^*\bar{D}^*(2S)[1/2(1/2^-)]$ states to $P_c(4312)$, $P_c(4440)$, and $P_c(4457)$ are of the same order of magnitude, around 1~MeV. Moreover, the dominant interaction mechanism in all three cases is the same, namely $\Sigma_c\bar{D}^*(2S)\to\Sigma_c\bar{D}^*+\pi$. For the decay into $P_c(4312)$, although the $\Sigma_c\bar{D}$ component plays a significant role in the $P_c(4312)$ wave function, the contribution from the process $\Sigma_c\bar{D}^*(2S)(1/2^-)\to\Sigma_c\bar{D}(1/2^-)+\pi$ is much smaller. This is because the spin-spin interaction in this channel is considerably weaker than that in $\Sigma_c\bar{D}^*(2S)(1/2^-)\to\Sigma_c\bar{D}^*(1/2^-,3/2^-)+\pi$.

In comparison with the above three cases, the decay width of the same initial molecular state to $P_c(4380)$ and a pion is also about 1~MeV. However, the underlying interaction mechanism is entirely different: the dominant contributions come from $\Sigma_c^*\bar{D}^*(2S)\to\Sigma_c^*\bar{D}^*+\pi$ and $\Sigma_c^*\bar{D}^*(2S)\to\Sigma_c^*\bar{D}+\pi$.

\subsection{$\Sigma_c\bar{D}^*(2S)/\Sigma_c^*\bar{D}(2S)/\Sigma_c^*\bar{D}^*(2S)[{1}/{2}({3}/{2}^-)] \to P_c+ \pi$}

\begin{figure}[!htbp]
\includegraphics[width=1\linewidth]{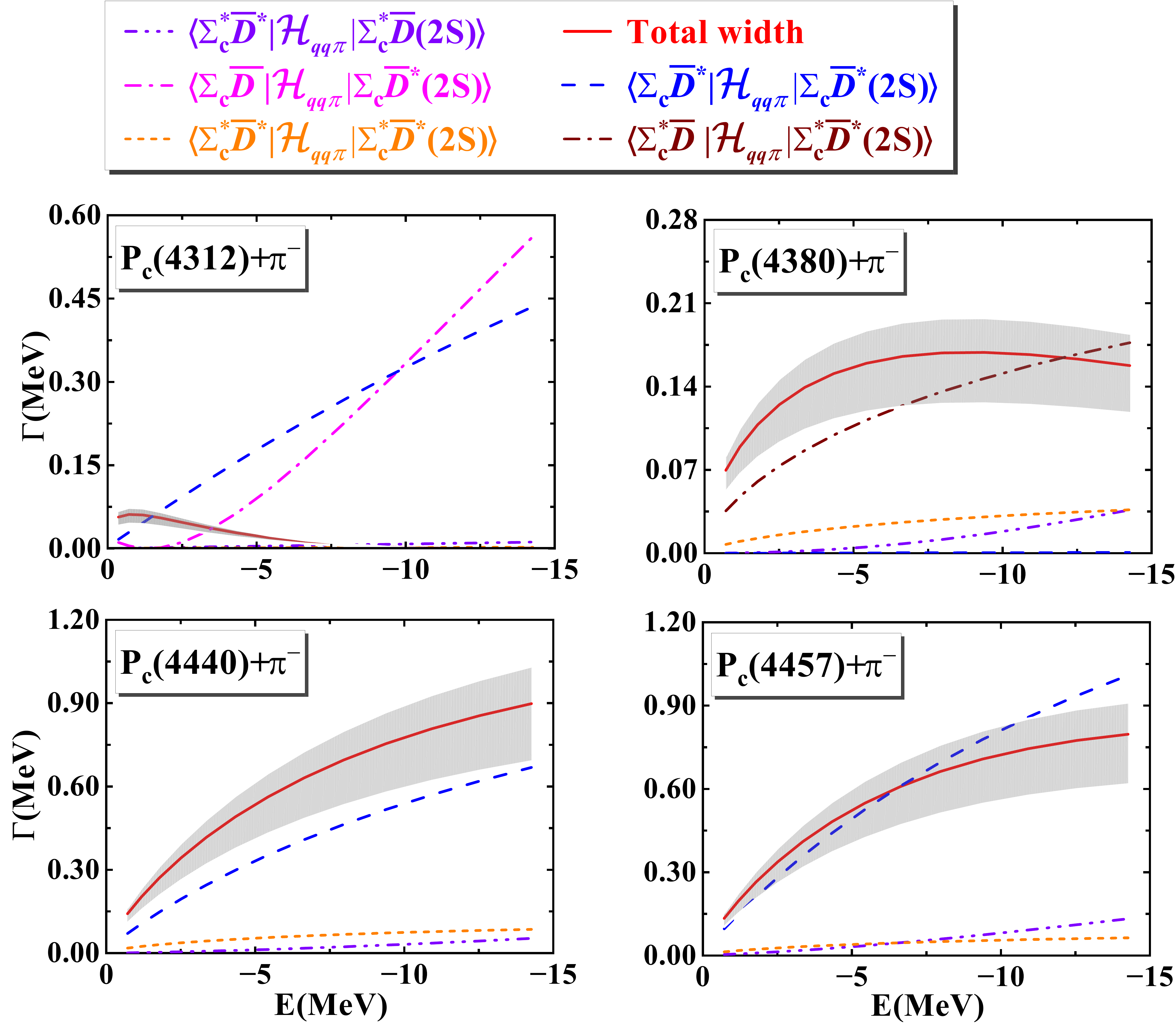}
\caption{The pion-emission widths for $\Sigma_c\bar{D}^*(2S)/\Sigma_c^*\bar{D}(2S)/\Sigma_c^*\bar{D}^*(2S)$ molecules with $I(J^P)=1/2(3/2^-)$ decaying into $P_c$ molecular states. Here, $E$ is the binding energy for the initial molecular states. $\langle A|\mathcal{H}_{qq\pi}|B\rangle$ corresponds to the pion-emission width for $B$ decaying to $A$. The cutoff values corresponding to the gray shaded area of the data are taken from 0.50 to 0.80 GeV, while the decay widths in the curve manner are calculated at $\Lambda=0.66$ GeV.}\label{fignum3}
\end{figure}

Compared with the $\Sigma_c\bar{D}^*(2S)/\Sigma_c^*\bar{D}^*(2S)[1/2(1/2^-)]$ case, as shown in Figure \ref{fignum3}, the pion-emission decay widths of $\Sigma_c\bar{D}^*(2S)/\Sigma_c^*\bar{D}(2S)/\Sigma_c^*\bar{D}^*(2S)[1/2(3/2^-)]$ to $P_c(4312)$, $P_c(4380)$, $P_c(4440)$, and $P_c(4457)$ are correspondingly smaller. The main reasons are as follows. For the final state $P_c(4312)$, the two dominant processes, $\Sigma_c\bar{D}^*(2S)\to\Sigma_c\bar{D}^*+\pi$ and $\Sigma_c\bar{D}^*(2S)\to\Sigma_c\bar{D}+\pi$, undergo strong destructive interference. For the final states $P_c(4440)$ and $P_c(4457)$, although the $\Sigma_c\bar{D}^*(2S)\to\Sigma_c\bar{D}^*+\pi$ interaction remains the largest contribution, as in the case with the initial spin-$1/2$ state, the spin-spin interaction in this process is weaker than that in the initial spin-$1/2$ case. Furthermore, the probabilities of the $\Sigma_c^*\bar{D}^*$ component in the initial state is much smaller than those in the $\Sigma_c\bar{D}^*(2S)/\Sigma_c^*\bar{D}^*(2S)[1/2(1/2^-)]$ state. This also explains why the pion-emission decay width to $P_c(4380)$ is smaller for the $\Sigma_c\bar{D}^*(2S)/\Sigma_c^*\bar{D}(2S)/\Sigma_c^*\bar{D}^*(2S)[1/2(3/2^-)]$ initial state.

\subsection{$\Sigma_c^*\bar{D}(2S)/\Sigma_c^*\bar{D}^*(2S)[{1}/{2}({3}/{2}^-)] \to P_c+ \pi$}

\begin{figure}[!htbp]
\includegraphics[width=1\linewidth]{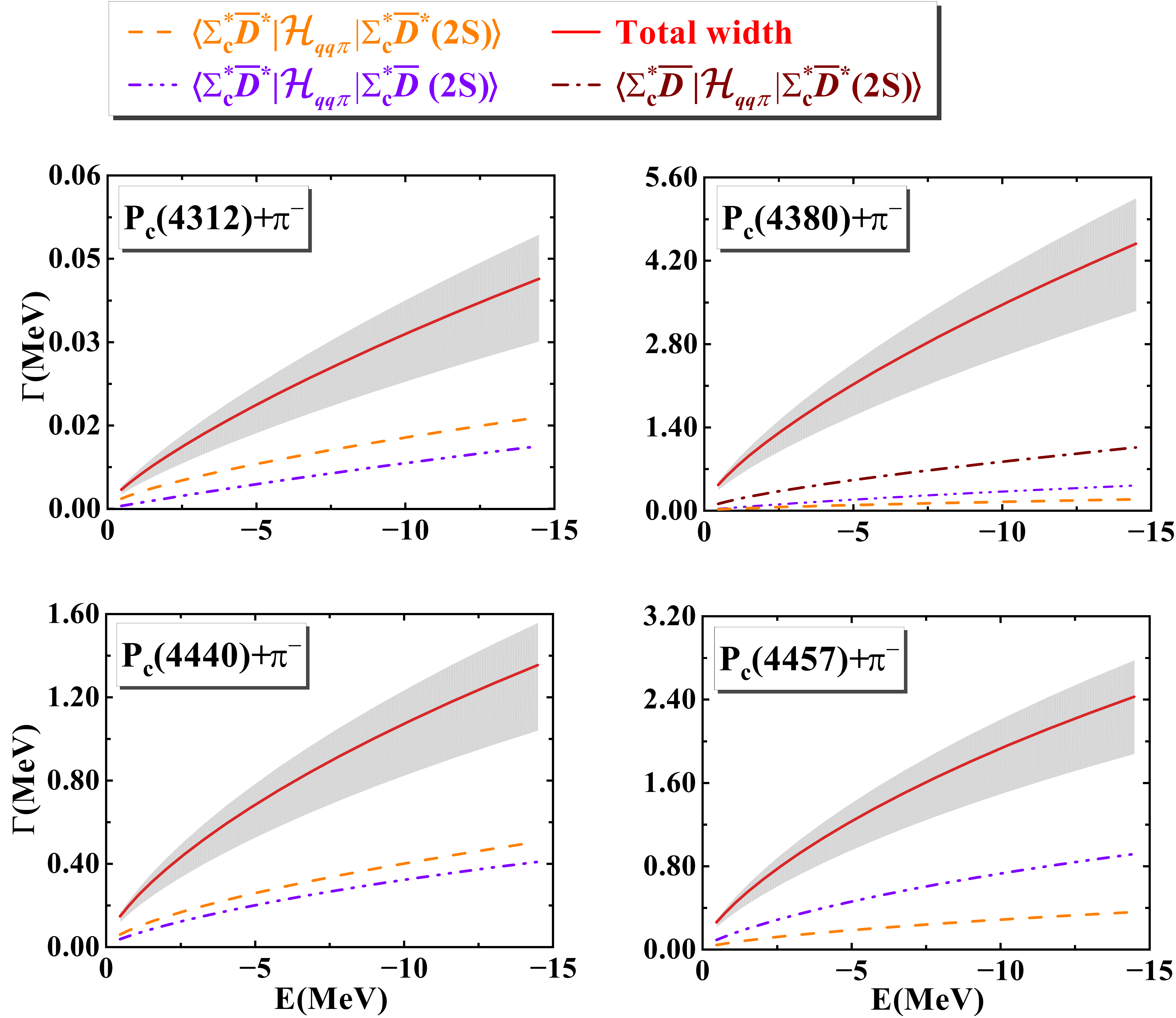}
\caption{The pion-emission widths for $\Sigma_c^*\bar{D}(2S)/\Sigma_c^*\bar{D}^*(2S)$ molecules with $I(J^P)=1/2(3/2^-)$ decaying into $P_c$ molecular states. Here, $E$ is the binding energy for the initial molecular states. $\langle A|\mathcal{H}_{qq\pi}|B\rangle$ corresponds to the pion-emission width for $B$ decaying to $A$. The cutoff values corresponding to the gray shaded area of the data are taken from 0.50 to 0.80 GeV, while the decay widths in the curve manner are calculated at $\Lambda=0.66$ GeV.}\label{fignum4}
\end{figure}

We then calculate the pion-emission decay widths of the $\Sigma_c^*\bar{D}(2S)/\Sigma_c^*\bar{D}^*(2S)[1/2(3/2^-)]$ state to $P_c(4312)$, $P_c(4380)$, $P_c(4440)$, and $P_c(4457)$. As shown in Figure \ref{fignum4}, for the final state $P_c(4312)$, only the processes $\Sigma_c^*\bar{D}(2S)\to\Sigma_c^*\bar{D}^*+\pi$ and $\Sigma_c^*\bar{D}^*(2S)\to\Sigma_c^*\bar{D}^*+\pi$ contribute the decay width at tree level. Since the probability of the $\Sigma_c^*\bar{D}^*$ component in the $P_c(4312)$ wave function is very small, the pion-emission width is extremely small — only a few tens of keV — when the binding energy of the initial molecular state satisfies $E > -15$~MeV.

For the final states $P_c(4380)$, $P_c(4440)$, and $P_c(4457)$, the pion-emission widths are all on the order of a few MeV. The dominant contributions arise from $\Sigma_c^*\bar{D}^*(2S)\to\Sigma_c^*\bar{D}+\pi$, $\Sigma_c^*\bar{D}^*(2S)\to\Sigma_c^*\bar{D}^*+\pi$, and $\Sigma_c^*\bar{D}(2S)\to\Sigma_c^*\bar{D}^*+\pi$, respectively. Because the probability of the $\Sigma_c^*\bar{D}$ component in $P_c(4380)$ is much larger than that of $\Sigma_c^*\bar{D}^*$, the pion-emission decay width of $\Sigma_c^*\bar{D}(2S)/\Sigma_c^*\bar{D}^*(2S)[1/2(3/2^-)]$ to $P_c(4380)$ is larger than that to $P_c(4440)$. Moreover, for a similar reason — namely, the $\Sigma_c^*\bar{D}(2S)$ component in the initial state has a larger probability than the $\Sigma_c^*\bar{D}^*(2S)$ component — the pion-emission width to $P_c(4457)$ is found to be larger than that to $P_c(4440)$.

\subsection{$\Sigma_c^*\bar{D}^*(2S)[{1}/{2}({1}/{2}^-, 3/2^-, 5/2^-)] \to P_c+ \pi$}

Finally, we also predict the pion-emission decay widths from the $S$-wave $\Sigma_c^*\bar{D}^*(2S)$ molecular states with $I=1/2$ to the ground-state $P_c$ molecular states. As shown in the Table \ref{tablenum}, we provide the pion-emission widths when the binding energy of the initial molecular states satisfies $E > -15$~MeV. Since the binding energy of the final $S$-wave $\Sigma_c^*\bar{D}^*$ molecular states with $I=1/2$ has not yet been determined, for simplicity, we vary the binding energies of both the initial and final states simultaneously.

\renewcommand\tabcolsep{0.2cm}
\renewcommand{\arraystretch}{1.7}
\begin{table}[htbp]
    \centering

    \caption{The $\pi^-$-emission widths for $\Sigma_c^*\bar{D}^*$ molecular states with $I(J^P)=1/2(1/2^-, 3/2^-, 5/2^-)$ decaying into $P_c$ molecular states. Here, $E$ stands for the the binding energy of the initial $\Sigma_c^*\bar{D}^*(2S)$ molecular states and the final $\Sigma_c^*\bar{D}^*$ molecules. The unites of binding energy $E$ and total decay width $\Gamma$ are MeV. \label{tablenum}}
    \begin{tabular}{c | c ccc}
        \toprule[1pt]\toprule[1pt] 
        \multirow{2}{*}{Initial States} & \multirow{2}{*}{Final States} & \multicolumn{3}{c}{{Total decay width $\Gamma$ }} \\
        &&$E=-5$     &$-10$    &$-15$ \\\midrule[1pt]
        
     $\Sigma_c^*\bar{D}^*(2S) \left[ \frac{1}{2} \left(\frac{1}{2}^-\right) \right] $ 
     &$P_c(4312)$    &$0.0016$ &$0.0043$  &$0.0075$  \\
     &$P_c(4380)$    &$0.3$  &$1.0$ &$2.1$ \\
     &$P_c(4440)$    &$0.05$ &$0.11$ &$0.18$  \\
     &$P_c(4457)$    &$0.48$  &$1.01$  &$1.62$ \\
        & $\Sigma_c^*\bar{D}^* \left[ \frac{1}{2} \left(\frac{1}{2}^-\right) \right] $ & $0.09$ & $0.26$  & $0.48$\\

        & $\Sigma_c^*\bar{D}^* \left[ \frac{1}{2} \left(\frac{3}{2}^-\right) \right] $ & $0.44$ & $1.34$  & $2.36$\\ 
        
        & $\Sigma_c^*\bar{D}^* \left[ \frac{1}{2} \left(\frac{5}{2}^-\right) \right] $ & $\sim0$   & $\sim0$   & $\sim0$  \\ 
        \midrule[1pt]
        
        $\Sigma_c^*\bar{D}^*(2S) \left[ \frac{1}{2} \left(\frac{3}{2}^-\right) \right] $ 
        &$P_c(4312)$    &$0.002$   &$0.006$ &$0.010$ \\
     &$P_c(4380)$    &$0.60$   &$1.58$ &$2.50$ \\
     &$P_c(4440)$    &$0.15$   &$0.30$ &$0.43$ \\
     &$P_c(4457)$    &$0.08$   &$0.17$ &$0.26$ \\
        & $\Sigma_c^*\bar{D}^* \left[ \frac{1}{2} \left(\frac{1}{2}^-\right) \right] $ & $0.20$  & $0.62$  & $1.18$\\

        & $\Sigma_c^*\bar{D}^* \left[ \frac{1}{2} \left(\frac{3}{2}^-\right) \right] $ & $0.03$ & $0.08$ & $0.15$\\ 
        
        & $\Sigma_c^*\bar{D}^* \left[ \frac{1}{2} \left(\frac{5}{2}^-\right) \right] $ & $0.03$ & $0.10$ & $0.21$ \\
        \midrule[1pt]
        
       $\Sigma_c^*\bar{D}^*(2S) \left[ \frac{1}{2} \left(\frac{5}{2}^-\right) \right] $ 
        &$P_c(4312)$    &$\sim0$  &$\sim0$  &$\sim0$ \\
     &$P_c(4380)$    &$0.018$   &$0.002$  &$0.023$ \\
     &$P_c(4440)$    &$\sim0$   &$\sim0$  &$\sim0$ \\
     &$P_c(4457)$    &$0.23$   &$0.54$  &$0.80$ \\
        & $\Sigma_c^*\bar{D}^* \left[ \frac{1}{2} \left(\frac{1}{2}^-\right) \right] $ & $\sim0$  & $\sim0$   & $\sim0$  \\
        & $\Sigma_c^*\bar{D}^* \left[ \frac{1}{2} \left(\frac{3}{2}^-\right) \right] $    & $0.16$  & $0.46$  & $0.98$ \\ 
        
        & $\Sigma_c^*\bar{D}^* \left[ \frac{1}{2} \left(\frac{5}{2}^-\right) \right] $ & $0.02$  & $0.04$ & $0.09$\\
        \bottomrule[1pt]\bottomrule[1pt] 
    \end{tabular}
    
\end{table}

Based on the numerical results in Table \ref{tablenum}, we can find that:
\begin{itemize}
    \item For the $\Sigma_c^*\bar{D}^*(2S)$ molecule with $I(J^P) = 1/2(1/2^-)$, the decay widths to $\Sigma_c^*\bar{D}^*$ molecule with $1/2(3/2^-)$, $P_c(4380)$ and $P_c(4457)$ can reach to a few MeV with $E>-15$ MeV, followed by $P_c(4440)$ and the $\Sigma_c^*\bar{D}^*$ molecule with $1/2(1/2^-)$, the widths to $P_c(4312)$ and the $\Sigma_c^*\bar{D}^*$ molecule with $1/2(5/2^-)$ are very tiny.
    \item For the $\Sigma_c^*\bar{D}^*(2S)$ molecule with $I(J^P) = 1/2(3/2^-)$, in the binding energy range of $E>-15$ MeV, the width to $P_c(4380)$ remains the largest, followed by the $\Sigma_c^*\bar{D}^*$ molecule with $1/2(1/2^-)$, the width to $P_c(4440)$, $P_c(4457)$, and the $\Sigma_c^*\bar{D}^*$ molecule with $1/2(3/2^-, 5/2^-)$ are on the same order of a few tenths of an MeV, the width to $P_c(4312)$ is still very small.
    \item For the $\Sigma_c^*\bar{D}^*(2S)$ molecule with $I(J^P) = 1/2(5/2^-)$, the width to $P_c(4380)$ drops sharply to $0.03$~MeV; the channel to $P_c(4457)$ and the $\Sigma_c^*\bar{D}^*$ molecule with $1/2(3/2^-)$ become relatively more prominent; the width to $P_c(4312)$, $P_c(4440)$, and the $\Sigma_c^*\bar{D}^*$ molecule with $1/2(1/2^-)$ are nearly zero.
    \item Overall, $P_c(4380)$ is the dominant decay recipient for low-spin excited states, while $P_c(4312)$ is strongly suppressed for all spins.
\end{itemize}


\section{Summary}\label{sec4}

In this work, we systematically study the pion-emission transitions from excited hidden-charm pentaquark molecules to the three known $P_c$ states within a hadronic molecular picture. We firstly extend our previous OBE potentials analysis to radially excited anti-charmed mesons, i.e., $\bar{D}(2S)$ and $\bar{D}^*(2S)$, and identify possible bound states of $\Sigma_c^{(*)}\bar{D}^{(*)}(2S)$. We then use the chiral quark model to calculate the decay widths in the rest frame of the initial molecule, where the effective Hamiltonian for the $\pi$-emission vertex is expanded up to $1/m^2$ corrections. The wave functions of both initial and final molecular states are obtained by solving the Schrödinger equation with OBE effective potentials, and coupled-channel effects among different $\Sigma_c^{(*)}\bar{D}^{(*)}$ components are also considered.

By numerical calculation, we can find: (1) For the $\Sigma_c\bar{D}(2S)/\Sigma_c\bar{D}^*(2S)/\Sigma_c^*\bar{D}^*(2S)[1/2(1/2^-)]$ initial state, the decay width to $P_c(4312)$ and $P_c(4440)$ can reach several MeV, while that to $P_c(4457)$ is below $0.3$~MeV due to destructive interference. The width to $P_c(4380)$ is about $0.25$~MeV, suppressed by the small $\Sigma_c^*\bar{D}^*(2S)$ component. (2) For the $\Sigma_c\bar{D}^*(2S)/\Sigma_c^*\bar{D}^*(2S)[1/2(1/2^-)]$, the widths to $P_c(4312)$, $P_c(4440)$ and $P_c(4457)$ are all around $1$~MeV, dominated by the $\Sigma_c\bar{D}^*(2S) \to \Sigma_c\bar{D}^*+\pi$ process. The width to $P_c(4380)$ is also $\sim 1$~MeV but originates from $\Sigma_c^*\bar{D}^*(2S)$ components. (3) For the $\Sigma_c\bar{D}^*(2S)/\Sigma_c^*\bar{D}(2S)/\Sigma_c^*\bar{D}^*(2S)$ molecule with $I(J^P)=1/2(3/2^-)$, the decay widths are generally smaller due to weaker spin-spin interactions and destructive interference. The width to $P_c(4312)$ is only a few tens of keV. (4) For the $\Sigma_c^*\bar{D}^*(2S)$ molecules with $I(J^P)=1/2(1/2^-, 3/2^-, 5/2^-)$, the decay patterns are distinct: $P_c(4380)$ dominates for low spins (width up to a few MeV), while for $J^P=5/2^-$ the $P_c(4457)$ channel becomes relatively prominent (up to $0.75$~MeV) and the $P_c(4312)$ channel is nearly zero. (5) According to the relations of the isospin wave functions, the neutral pion emission widths are almost half that of the charged pion-emission widths.

It is instructive to compare our results with the earlier studies of pionic transitions among ground-state $P_c$ molecules \cite{Xiao:2020frg,Ling:2021lmq}. Those works focused on decays such as $P_c(4457)\to P_c(4312)\pi$, while here we consider the decay of radially excited molecules built from $\bar{D}^{(*)}(2S)$ into the ground-state $P_c$ states. Despite the different initial states, the spin-dependent selection rules and interference patterns are qualitatively consistent: both frameworks predict strong suppression of certain channels due to destructive interference (e.g., the $P_c(4457)$ mode for the $I(J^P)=1/2(1/2^-)$ initial state). Quantitatively, our widths (a few MeV) are comparable to those found in Refs. \cite{Xiao:2020frg,Ling:2021lmq} after accounting for phase-space differences, which supports the overall consistency of the molecular picture.

In summary, the pion-emission decays provide a powerful probe to distinguish different spin-parity assignments of excited molecular pentaquarks. The predicted widths can be within the reach of current and future experiments. This work establishes for the first time a decay connection between radially excited anti-charmed meson-based molecules and the observed $P_c$ states, opening a new avenue for understanding the excited spectrum of hidden-charm pentaquarks. We hope facilities like the upgraded LHCb and the future PANDA experiment can detect such decay processes. If signals of excited $P_c$ molecular states decaying via $\pi$ emission to the known $P_c$ states could be observed, it would provide decisive evidence for the molecular interpretation of pentaquarks and simultaneously open a new chapter in exotic hadron spectroscopy.  

\section*{ACKNOWLEDGMENTS}

We would like to thank Hui-Hua Zhong, Ru-Hui Ni and Xian-Hui Zhong for their valuable discussions. This project is supported by the National Natural Science Foundation of China under Grant No. 12305139, and the Xiaoxiang Scholars Programme of Hunan Normal University. F. L. Wang is also supported by the National Natural Science Foundation of China under Grants No. 12335001 and No. 12405097.

\appendix

\section{The OBE effective potentials and bound states properties for possible $\Sigma_c^{(*)}\bar{D}^{(*)}(2S)$ molecules}\label{app01}

The general procedures for deducing the OBE effective potentials can be devided as follows: after constructed the effective Lagrangians, we can write down the scattering amplitudes for $\Sigma_c^{(*)}\bar{D}^{(*)}(2S)\to \Sigma_c^{(*)}\bar{D}^{(*)}(2S)$ processes in t-channel by exchanging one light meson. Next, we can derive the OBE effective potentials for the discussed channels in momentum space based on the Breit approximation, i.e.,
\begin{eqnarray}
  \mathcal{V}_{E}(q) = -\frac{\mathcal{M}(\Sigma_c^{(*)}\bar{D}^{(*)}(2S)\to \Sigma_c^{(*)}\bar{D}^{(*)}(2S))}{\sqrt{2m_{\Sigma_c^{(*)}}^i 2m_{\Sigma_c^{(*)}}^f 2m_{\bar{D}^{(*)}(2S)}^i 2m_{\bar{D}^{(*)}(2S)}^f}}.
  \end{eqnarray}
Here, notations $i$ and $f$ label the initial and final states, respectively. $E$ stands for the exchanged mesons, including the scalar meson $\sigma$, the pseudoscalar mesons $\pi/\eta$, the vector mesons $\rho/\omega$. With the help of the Fourier transform, we can finally obtain the OBE effective potentials in the coordinate space $\mathcal{V}(r)$
\begin{eqnarray}
\mathcal{V}(r) = \int \frac{d^3\vec{q}}{(2\pi)^{3}}e^{i\vec{q}\cdot r}\mathcal{V}_{E}(q)\mathcal{F}^2(q^2,m_E^2).
\end{eqnarray}
Here, we introduce a monopole form factor, $\mathcal{F}(q^2,m_E^2)=(\Lambda^2-m_E^2)/(\Lambda^2-q^2)$, at every interaction vertexes. This from factor can compensate the off-shell effects of the exchanged particles. $\Lambda$, $m_E$, and $q$ stand for the cutoff, the mass, and the four momentum of the exchanged mesons. The reasonable value of the cutoff $\Lambda$ are in the range of $\Lambda\sim1.00$ GeV, based on the experience of the study of deuteron~\cite{Tornqvist:1993ng,Tornqvist:1993vu}.

\renewcommand\tabcolsep{0.35cm}
\renewcommand{\arraystretch}{1.6}
\begin{table*}[!htbp]
  \centering
  \caption{The OBE effective potentials for all scattering processes with $I=1/2$. Here, we define several functions, $Y_{\Lambda,m} = \frac{1}{4\pi r}(e^{-mr} - e^{-\Lambda r}) - \frac{\Lambda^2-m^2}{8\pi\Lambda}e^{-\Lambda r}$, $\mathcal{Y}^{ij}_{\Lambda, m_a} = \mathcal{D}_{ij}Y_{\Lambda,m}$, $\mathcal{Z}^{ij}_{\Lambda m_a} = \left(\mathcal{E}_{ij}\nabla^2 + \mathcal{F}_{ij}r\frac{\partial}{\partial r}\frac{1}{r}\frac{\partial}{\partial r}\right) Y_{\Lambda,m}$, and $\mathcal{Z}^{\prime ij}_{\Lambda m_a} = \left(2\mathcal{E}_{ij}\nabla^2 - \mathcal{F}_{ij}r\frac{\partial}{\partial r}\frac{1}{r}\frac{\partial}{\partial r}\right) Y_{\Lambda,m}$. $\mathcal{D}_{ij}$, $\mathcal{E}_{ij}$, and $\mathcal{F}_{ij}$ denote the spin-spin interaction and tensor operators, respectively. The variables in these functions are defined as $\Lambda_i^2 = \Lambda^2 - q_i^2$, $m_i^2 = m^2 - q_i^2$, with $i = 1,\dots,5$. And $q_0^2 = \left(\frac{M_{\Sigma_c^*}^2+M_{\bar{D}^*(2S)}^2-M_{\Sigma_c}^2-M_{\bar{D}(2S)}^2}{2(M_{\Sigma_c}+M_{\bar{D}^*(2S)})}\right)^2$, $q_1^2 = \left(\frac{M_{\bar{D}^*(2S)}^2-M_{\bar{D}(2S)}^2}{2(M_{\Sigma_c}+M_{\bar{D}^*(2S)})}\right)^2$, $q_2^2 = \left(\frac{M_{\Sigma_c^*}^2-M_{\Sigma_c}^2}{2(M_{\Sigma_c^*}+M_{\bar{D}^*(2S)})}\right)^2$, $q_3^2 = \left(\frac{M_{\Sigma_c^*}^2-M_{\Sigma_c}^2}{2(M_{\Sigma_c^*}+M_{\bar{D}(2S)})}\right)^2$, $q_4^2 = \left(\frac{M_{\bar{D}^*(2S)}^2-M_{\bar{D}(2S)}^2}{2(M_{\Sigma_c}+M_{\bar{D}(2S)})}\right)^2$, $q_5^2 = \left(\frac{M_{\Sigma_c}^2+M_{\bar{D}^*(2S)}^2-M_{\Sigma_c^*}^2-M_{\bar{D}(2S)}^2}{2(M_{\Sigma_c^*}+M_{\bar{D}^*(2S)})}\right)^2$.}
  \label{potentials}

  \begin{tabular}{c | l | c | l} 
    \hline \hline
    Processes & \multicolumn{1}{c|}{OBE effective potentials} & Processes & \multicolumn{1}{c}{OBE effective potentials} \\
    \hline
    $\Sigma_c \bar{D}(2S)\to\Sigma_c \bar{D}(2S)$ &
    $\begin{aligned}[t]
       &-l_S g_S \mathcal{Y}^{11}_{\Lambda, m_{\sigma}} - \frac{\mathcal{G}}{2}\beta\beta_S g_V^2 \mathcal{Y}^{11}_{\Lambda, m_{\rho}} \\
       &- \frac{1}{4}\beta\beta_S g_V^2 \mathcal{Y}^{11}_{\Lambda, m_{\omega}}
    \end{aligned}$ &
    $\Sigma_c \bar{D}(2S)\to\Sigma_c^* \bar{D}(2S)$ &
    $\begin{aligned}[t]
       &\frac{\mathcal{G}\beta\beta_Sg_V^2}{2\sqrt{3}}\mathcal{Y}^{12}_{\Lambda_3,m_{\rho3}}  +\frac{\beta\beta_Sg_V^2}{4\sqrt{3}}\mathcal{Y}^{12}_{\Lambda_3,m_{\omega3}}
    \end{aligned}$ \\
    \hline
    
    $\Sigma_c \bar{D}(2S)\to\Sigma_c \bar{D}^*(2S)$ &
    $\begin{aligned}[t]
       &\frac{\mathcal{G}gg_1}{3f_\pi^2}\mathcal{Z}^{13}_{\Lambda_4,m_{\pi4}} +\frac{gg_1}{18f_\pi^2}\mathcal{Z}^{13}_{\Lambda_4,m_{\eta4}} \\
       &+\frac{2\mathcal{G}\lambda\lambda_Sg_V^2}{9}\mathcal{Z}^{\prime13}_{\Lambda_4,m_{\rho4}} +\frac{\lambda\lambda_Sg_V^2}{9}\mathcal{Z}^{\prime13}_{\Lambda_4,m_{\omega4}}
    \end{aligned}$ &
    $\Sigma_c \bar{D}(2S)\to\Sigma_c^* \bar{D}^*(2S)$ &
    $\begin{aligned}[t]
       &\frac{\mathcal{G}gg_1}{2\sqrt{3}f_\pi^2}\mathcal{Z}^{14}_{\Lambda_5,m_{\pi5}} +\frac{{gg_1}}{12\sqrt{3}{f_\pi^2}}\mathcal{Z}^{14}_{\Lambda_5,m_{\eta5}} \\
       &+\frac{\mathcal{G}\lambda\lambda_Sg_V^2}{3\sqrt{3}}\mathcal{Z}^{\prime14}_{\Lambda_5,m_{\rho5}} +\frac{\lambda\lambda_Sg_V^2}{6\sqrt{3}}\mathcal{Z}^{\prime14}_{\Lambda_5,m_{\omega5}}
    \end{aligned}$ \\
    \hline
     
    $\Sigma_c^* \bar{D}(2S)\to\Sigma_c^* \bar{D}(2S)$ &
    $\begin{aligned}[t]
       &-l_Sg_S\mathcal{Y}^{22}_{\Lambda,m_{\sigma}} -\frac{\mathcal{G}\beta\beta_Sg_V^2}{2}\mathcal{Y}^{22}_{\Lambda,m_{\rho}} \\
       &-\frac{\beta\beta_Sg_V^2}{4}\mathcal{Y}^{22}_{\Lambda,m_{\omega}}
    \end{aligned}$ &
    $\Sigma_c^* \bar{D}(2S)\to\Sigma_c \bar{D}^*(2S)$ &
    $\begin{aligned}[t]
       &\frac{\mathcal{G}{gg_1}}{2\sqrt{3}{f_\pi^2}}\mathcal{Z}^{23}_{\Lambda_0,m_{\pi0}} +\frac{{gg_1}}{12\sqrt{3}{f_\pi^2}}\mathcal{Z}^{23}_{\Lambda_0,m_{\eta0}} \\
       &-\frac{\mathcal{G}\lambda\lambda_Sg_V^2}{3\sqrt{3}}\mathcal{Z}^{\prime23}_{\Lambda_0,m_{\rho0}} -\frac{\lambda\lambda_Sg_V^2}{6\sqrt{3}}\mathcal{Z}^{\prime23}_{\Lambda_0,m_{\omega0}}
    \end{aligned}$ \\
    \hline
    
    $\Sigma_c^* \bar{D}(2S)\to\Sigma_c^* \bar{D}^*(2S)$ &
    $\begin{aligned}[t]
       &\frac{\mathcal{G}{gg_1}}{2{f_\pi^2}}\mathcal{Z}^{24}_{\Lambda_1,m_{\pi1}} +\frac{{gg_1}}{12{f_\pi^2}}\mathcal{Z}^{24}_{\Lambda_1,m_{\eta1}} \\
       &-\frac{\mathcal{G}\lambda\lambda_Sg_V^2}{3}\mathcal{Z}^{\prime24}_{\Lambda_1,m_{\rho1}} -\frac{\lambda\lambda_Sg_V^2}{6}\mathcal{Z}^{\prime 24}_{\Lambda_1,m_{\omega1}}
    \end{aligned}$ &
    $\Sigma_c \bar{D}^*(2S)\to\Sigma_c \bar{D}^*(2S)$ &
    $\begin{aligned}[t]
       &-l_Sg_S\mathcal{Y}^{33}_{\Lambda,m_{\sigma}} +\frac{\mathcal{G}{gg_1}}{3{f_\pi^2}}\mathcal{Z}^{33}_{\Lambda,m_{\pi}} \\
       &+\frac{{gg_1}}{18{f_\pi^2}}\mathcal{Z}^{33}_{\Lambda,m_{\eta}} -\frac{\mathcal{G}\beta\beta_Sg_V^2}{2}\mathcal{Y}^{33}_{\Lambda,m_{\rho}} \\
       &-\frac{2\mathcal{G}\lambda\lambda_Sg_V^2}{9}\mathcal{Z}^{\prime33}_{\Lambda,m_{\rho}} -\frac{\beta\beta_Sg_V^2}{4}\mathcal{Y}^{33}_{\Lambda,m_{\omega}} \\
       &-\frac{\lambda\lambda_Sg_V^2}{9}\mathcal{Z}^{\prime33}_{\Lambda,m_{\omega}}
    \end{aligned}$ \\
    \hline
    
    $\Sigma_c \bar{D}^*(2S)\to\Sigma_c^* \bar{D}^*(2S)$ &
    $\begin{aligned}[t]
       &\frac{l_Sg_S}{\sqrt{3}}\mathcal{Y}^{34}_{\Lambda_2,m_{\sigma2}} +\frac{\sqrt{3}\mathcal{G}{gg_1}}{6{f_\pi^2}}\mathcal{Z}^{34}_{\Lambda_2,m_{\pi2}} \\
       &+\frac{\sqrt{3}{gg_1}}{36{f_\pi^2}}\mathcal{Z}^{34}_{\Lambda_2,m_{\eta2}} +\frac{\mathcal{G}\beta\beta_Sg_V^2}{2\sqrt{3}}\mathcal{Y}^{34}_{\Lambda_2,m_{\rho2}} \\
       &-\frac{\mathcal{G}\lambda\lambda_Sg_V^2}{3\sqrt{3}}\mathcal{Z}^{\prime34}_{\Lambda_2,m_{\rho2}} +\frac{\beta\beta_Sg_V^2}{4\sqrt{3}}\mathcal{Y}^{34}_{\Lambda_2,m_{\omega2}} \\
       &-\frac{\lambda\lambda_Sg_V^2}{6\sqrt{3}}\mathcal{Z}^{\prime 34}_{\Lambda_2,m_{\omega2}}
    \end{aligned}$ &
    $\Sigma_c^* \bar{D}^*(2S)\to\Sigma_c^* \bar{D}^*(2S)$ &
    $\begin{aligned}[t]
       &-l_Sg_S\mathcal{Y}^{44}_{\Lambda,m_{\sigma}} -\frac{\mathcal{G}{gg_1}}{2{f_\pi^2}}\mathcal{Z}^{44}_{\Lambda,m_{\pi}} \\
       &-\frac{{gg_1}}{12{f_\pi^2}}\mathcal{Z}^{44}_{\Lambda,m_{\eta}} -\frac{\mathcal{G}\beta\beta_Sg_V^2}{2}\mathcal{Y}^{44}_{\Lambda,m_{\rho}} \\
       &+\frac{\mathcal{G}\lambda\lambda_Sg_V^2}{3}\mathcal{Z}^{\prime44}_{\Lambda,m_{\rho}} -\frac{\beta\beta_Sg_V^2}{4}\mathcal{Y}^{44}_{\Lambda,m_{\omega}} \\
       &+\frac{\lambda\lambda_Sg_V^2}{6}\mathcal{Z}^{\prime44}_{\Lambda,m_{\omega}}
    \end{aligned}$ \\
    \hline \hline
  \end{tabular}
\end{table*}

\begin{figure*}[!htbp]
\includegraphics[width=1\linewidth]{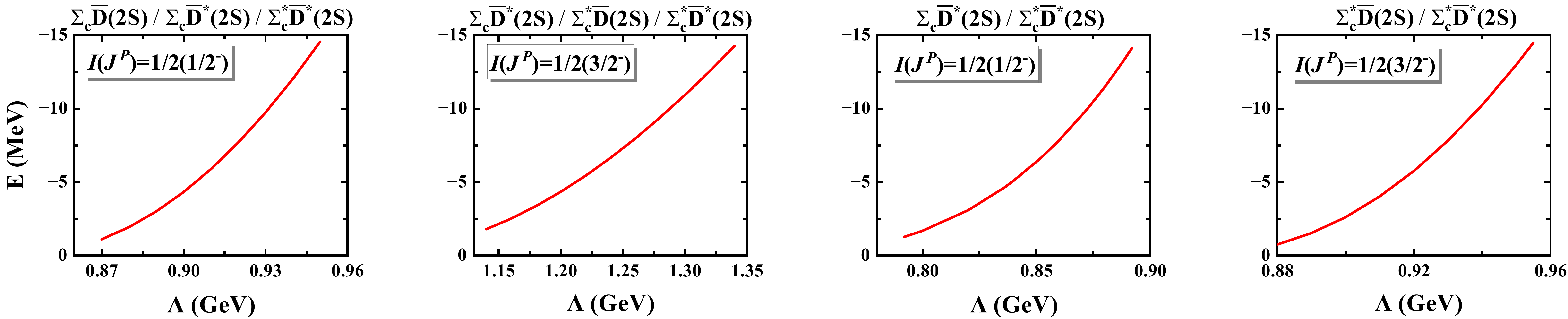}
\includegraphics[width=1\linewidth]{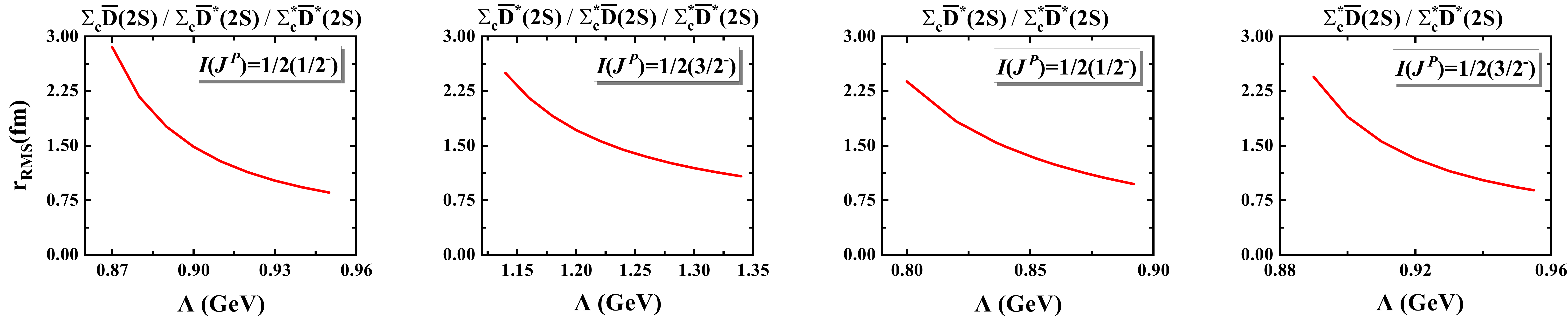}
\caption{The cutoff dependence of the loosely bound states solutions (the binding energy $E$ and the root-mean-square radii $r_{\text{RMS}}$) for the coupled $\Sigma_c^{(*)}\bar{D}^{(*)}(2S)$ states.}\label{binding}
\end{figure*}

If considering the heavy quark limit and chiral symmetry \cite{Yan:1992gz,Wise:1992hn,Burdman:1992gh,Casalbuoni:1996pg,Falk:1992cx,Liu:2011xc}, the relevant effective Lagrangians are
\begin{eqnarray}
\mathcal{L}_{H} &=& g_S\langle \bar{H}_a^{(\bar{Q})}\sigma H_b^{(\bar{Q})}\rangle
  +ig\langle \bar{H}_a^{(\bar{Q})}\gamma_{\mu}A_{ab}^{\mu}\gamma_5H_b^{(\bar{Q})}\rangle\nonumber\\
  &&-i\beta\langle \bar{H}_a^{(\bar{Q})}v_{\mu}\left(\mathcal{V}_{ab}^{\mu}-\rho_{ab}^{\mu}\right)
  H_b^{(\bar{Q})}\rangle\nonumber\\
  &&+i\lambda\langle \bar{H}_a^{(\bar{Q})}\sigma_{\mu\nu}F^{\mu\nu}(\rho)H_b^{(\bar{Q})}\rangle,\label{lag1}\\
\mathcal{L}_{\mathcal{S}} &=& l_S\langle\bar{\mathcal{S}}_{\mu}\sigma\mathcal{S}^{\mu}\rangle -\frac{3}{2}g_1\varepsilon^{\mu\nu\lambda\kappa}v_{\kappa}\langle\bar{\mathcal{S}}_{\mu}
A_{\nu}\mathcal{S}_{\lambda}\rangle\nonumber\\
&&+i\beta_{S}\langle\bar{\mathcal{S}}_{\mu}v_{\alpha}\left(\mathcal{V}_{ab}^{\alpha}
-\rho_{ab}^{\alpha}\right) \mathcal{S}^{\mu}\rangle
+\lambda_S\langle\bar{\mathcal{S}}_{\mu}F^{\mu\nu}(\rho)\mathcal{S}_{\nu}\rangle,\nonumber\\
\end{eqnarray}
where the multiplet field $H^{(\bar{Q})}$ is composed by the pseudoscalar meson $\tilde{\mathcal{P}}=\left(\bar{D}^0,\,D^-\right)^T$ and vector meson $\tilde{\mathcal{P}}^*=\left(\bar{D}^{*0},\,D^{*-}\right)^T$, its conjugate field of $\bar{H}^{(\bar{Q})}$ satisfies $\bar{H}^{(\bar{Q})}=\gamma_0H^{(\bar{Q})\dag}\gamma_0$. $\mathcal{S}$ is defined as a superfield, which includes $\mathcal{B}_6$ with $J^P=1/2^+$ and $\mathcal{B}^*_6$ with $J^P=3/2^+$ in the $6_F$ flavor representation. Here, their expressions read as
\begin{eqnarray}
H^{(\bar{Q})} &=& [\tilde{\mathcal{P}}^{*\mu}\gamma_{\mu}-\tilde{\mathcal{P}}\gamma_5]\frac{1-\rlap\slash v}{2},\\
\mathcal{S}_{\mu} &=&
-\sqrt{\frac{1}{3}}(\gamma_{\mu}+v_{\mu})\gamma^5\mathcal{B}_6
       +\mathcal{B}_{6\mu}^*,
\end{eqnarray}
with
\begin{eqnarray}
\mathcal{B}_6^{(*)} = \left(\begin{array}{cc}
         \Sigma_c^{{(*)}++}              &\frac{1}{\sqrt{2}}\Sigma_c^{{(*)}+}\\
         \frac{1}{\sqrt{2}}\Sigma_c^{{(*)}+}      &\Sigma_c^{{(*)}0}
\end{array}\right).
\end{eqnarray}
The expressions of axial current and vector current denote as
\begin{eqnarray*}
A_{\mu} &=& \frac{1}{2}(\xi^{\dag}\partial_{\mu}\xi-\xi\partial_{\mu}\xi^{\dag})=\frac{i}{f_{\pi}}
\partial_{\mu}\mathbb{P}+\ldots,\\
\mathcal{V}_{\mu} &=&
\frac{1}{2}(\xi^{\dag}\partial_{\mu}\xi-\xi\partial_{\mu}\xi^{\dag})
=\frac{i}{2f_{\pi}^2}\left[\mathbb{P},\partial_{\mu}\mathbb{P}\right]+\ldots,
\end{eqnarray*}
respectively, with $\xi=\text{exp}(i\mathbb{P}/f_{\pi})$ and the pion decay constant $f_{\pi}=132$ MeV. $\rho_{ba}^{\mu}=ig_V\mathbb{V}_{ba}^{\mu}/\sqrt{2}$, $F^{\mu\nu}(\rho)=\partial^{\mu}\rho^{\nu}-\partial^{\nu}\rho^{\mu}
+\left[\rho^{\mu},\rho^{\nu}\right]$. $\mathbb{P}$ and $\mathbb{V}$ stand for the isoscalar and vector matrixes, respectively, i.e.,
\begin{eqnarray*}\left.\begin{array}{ll}
\mathbb{P} = \left(\begin{array}{cc}
\frac{\pi^0}{\sqrt{2}}+\frac{\eta}{\sqrt{6}} &\pi^+\\
\pi^- &-\frac{\pi^0}{\sqrt{2}}+\frac{\eta}{\sqrt{6}}
\end{array}\right),
&\mathbb{V} = \left(\begin{array}{cc}
\frac{\rho^0}{\sqrt{2}}+\frac{\omega}{\sqrt{2}}  &\rho^+\\
\rho^- &-\frac{\rho^0}{\sqrt{2}}+\frac{\omega}{\sqrt{2}}
\end{array}\right).
\end{array}\right.
\end{eqnarray*}

\renewcommand\tabcolsep{0.12cm}
\renewcommand{\arraystretch}{1.7}
\begin{table}[!htbp]
\caption{The coupling constants and the corresponding phase factors adopted in this work. The masses for the involvd particles are in the unites of MeV.\label{coupling}}
\begin{tabular}{cccccc}
\hline \hline
 $g_s$    &$g$     &$\beta$      &$\lambda~(\text{GeV}^{-1})$    &$g_V$   &$f_{\pi}~(\text{GeV})$\\
 0.76     &0.59    &0.9          &0.56    &5.8     &$0.132$\\\hline
 $l_B$    &$\beta_B$          &$l_s$    &$g_1$        &$\beta_S$    &$\lambda_S~(\text{GeV}^{-1})$\\
 $-3.1$   &$-0.87$    &6.2      &0.94         &1.74         &3.31\\\hline
 $m_{\Sigma_c}$  &$m_{\Sigma_c^{*}}$  &$m_{\bar{D}}$ &$m_{\bar{D^*}}$    &$m_{\bar{D}(2S)}$  &$m_{\bar{D^*}(2S)}$\\
 $2453.54$   &$2518.10$    &1867.21    &2008.61      &2550.00     &2600.00 \\\hline
 $m_{\pi^{\pm}}$    &$m_{\pi}$ &$m_{\sigma}$ &$m_{\rho}$  &$m_{\omega}$   &$m_{\eta}$    \\
 $139.57$   &$137.27$  &600.00      &775.49         &782.65      &547.85\\
 \hline \hline
\end{tabular}
\end{table}

In Table \ref{coupling}, we collect the relevant coupling constants and the masses for all involved particles. Finally, we can derive the OBE effective potentials for the coupled $\Sigma_c^{(*)}\bar{D}^{(*)}(2S)$ systems. In Table \ref{potentials} and Table \ref{matrix}, we summary the corresponding OBE effective potentials and matrices elements for the spin-spin interactions and tensor force operators.

We next solve the coupled channel Schr\"{o}dinger equations with these OBE effective potentials. As shown in Figure \ref{binding} and Table \ref{tab:result}, we can obtain loosely bound states solutions for the coupled $\Sigma_c^{(*)}\bar{D}^{(*)}(2S)$ systems. Here, the cutoff $\Lambda$ for each coupled-channel system is treated as a phenomenological parameter; we vary it uniformly across all exchanged mesons within a given channel to produce bound states with binding energies $E>-15$ MeV. Table \ref{tab:result} lists the cutoff values for each molecular candidate. It is observed that the $J^P = 3/2^-$ state requires $\Lambda \sim 1.21$--$1.34$~GeV, larger than 1~GeV. This is physically due to the weaker tensor and spin-spin attractions in this channel, which necessitate a stronger short-range force to bind the system. Such a behavior is consistent with findings in other hadronic molecular calculations  \cite{Chen:2015loa,Chen:2019asm,Tornqvist:1993ng,Tornqvist:1993vu}.

In summary, our calculations suggest the following systems as possible molecular candidates: The $\Sigma_c\bar{D}(2S)/\Sigma_c\bar{D}^*(2S)/\Sigma_c^*\bar{D}^*(2S)$ molecule with $I(J^P)=1/2(1/2^-)$, the $\Sigma_c\bar{D}^*(2S)/\Sigma_c^*\bar{D}(2S)/\Sigma_c^*\bar{D}^*(2S)$ molecule with $I(J^P)=1/2(1/2^-, 3/2^-)$, the $\Sigma_c^*\bar{D}(2S)/\Sigma_c^*\bar{D}^*(2S)$ molecule with $I(J^P)=1/2(3/2^-)$, and the $\Sigma_c^*\bar{D}^*(2S)$ molecule with $I(J^P)=1/2(1/2^-, 3/2^-, 5/2^-)$.

\renewcommand\tabcolsep{0.35cm}
\renewcommand{\arraystretch}{1.25}
\begin{table*}[!htbp]
\centering
\caption{Matrix elements $\langle f|\Omega|i\rangle$ for the spin-spin interactions and tensor force operators in the OBE effective potentials. Here, we have $\langle f|\mathcal{E}_{32}|i\rangle^T = \langle f|\mathcal{E}_{23}|i\rangle$, $\langle f|\mathcal{F}_{32}|i\rangle^T = \langle f|\mathcal{F}_{23}|i\rangle$, $\langle f|\mathcal{E}_{42}|i\rangle^T = \langle f|\mathcal{E}_{24}|i\rangle$, $\langle f|\mathcal{F}_{42}|i\rangle^T = \langle f|\mathcal{F}_{24}|i\rangle$, $\langle f|\mathcal{E}_{43}|i\rangle^T = \langle f|\mathcal{E}_{34}|i\rangle$, $\langle f|\mathcal{F}_{43}|i\rangle^T = \langle f|\mathcal{F}_{34}|i\rangle$ and $\langle f|\mathcal{D}_{43}|i\rangle^T = \langle f|\mathcal{D}_{34}|i\rangle = (0)$.}\label{matrix}
\resizebox{\textwidth}{!}{
\begin{tabular}{c|c|c|c}
\hline \hline
$\langle f|\Omega|i\rangle$ & $1/2^-$ & $3/2^-$ & $5/2^-$ \\ \hline
$\langle\Sigma_c^* \bar{D}(2S){|\mathcal{D}_{22}|}\Sigma_c^* \bar{D}(2S)\rangle$ & 
$\begin{pmatrix} 1 \end{pmatrix}$ & 
$\begin{pmatrix} 1 & 0 \\ 0 & 1 \end{pmatrix}$ & 
$\begin{pmatrix} 1 \end{pmatrix}$ \\ \hline
$\langle\Sigma_c^* \bar{D}(2S){|\mathcal{E}_{23}|}\Sigma_c \bar{D}^*(2S)\rangle$ & 
$\begin{pmatrix} 0 & 1 \end{pmatrix}$ & 
$\begin{pmatrix} 1 & 0 & 0 \\0 & 0 & 1 \end{pmatrix}$ & 
$\begin{pmatrix} 0 & 1 \end{pmatrix}$ \\ 
$\langle\Sigma_c^* \bar{D}(2S){|\mathcal{F}_{23}|}\Sigma_c \bar{D}^*(2S)\rangle$ & 
$\begin{pmatrix} -\sqrt{2} & 1 \end{pmatrix}$ & 
$\begin{pmatrix} 0 & 1 & -1 \\ -1 & -1 & 0 \end{pmatrix}$ & 
$\begin{pmatrix} \sqrt{\frac{2}{7}} & -\frac{7}{5} \end{pmatrix}$ \\ \hline
$\langle\Sigma_c^* \bar{D}(2S){|\mathcal{E}_{24}|}\Sigma_c^* \bar{D}^*(2S)\rangle$ & 
$\begin{pmatrix} 0 & \sqrt{\frac{5}{3}} & 0 \end{pmatrix}$ & 
$\begin{pmatrix} \sqrt{\frac{5}{3}}  & 0 & 0 & 0 \\ 0  & 0 & \sqrt{\frac{5}{3}}  & 0 \end{pmatrix}$ & 
$\begin{pmatrix} 0 & 0 & \sqrt{\frac{5}{3}} & 0 \end{pmatrix}$ \\ 
$\langle\Sigma_c^* \bar{D}(2S){|\mathcal{F}_{24}|}\Sigma_c^* \bar{D}^*(2S)\rangle$ & 
$\begin{pmatrix} \frac{1}{\sqrt{3}} & -\frac{4}{\sqrt{15}} & -\sqrt{\frac{3}{5}} \end{pmatrix}$ & 
$\begin{pmatrix} 0 & -\frac{1}{\sqrt{6}} & \frac{4}{\sqrt{15}} &  -\sqrt{\frac{21}{10}} \\ \frac{4}{\sqrt{15}} & \frac{1}{\sqrt{6}} & 0 & -\sqrt{\frac{15}{14}} \end{pmatrix}$ & 
$\begin{pmatrix} \sqrt{\frac{7}{5}} & -\frac{1}{\sqrt{21}} & \frac{4\sqrt{\frac{5}{3}}}{7} &  -\frac{\sqrt{10}}{7} \end{pmatrix}$ \\ \hline
$\langle\Sigma_c \bar{D}^*(2S){|\mathcal{D}_{33}|}\Sigma_c \bar{D}^*(2S)\rangle$ & 
$\begin{pmatrix} 1 & 0 \\ 0 & 1 \end{pmatrix}$ & 
$\begin{pmatrix} 1 & 0 & 0 \\ 0 & 1 & 0 \\ 0 & 0 & 1 \end{pmatrix}$ & 
$\begin{pmatrix} 1 & 0 \\ 0 & 1 \end{pmatrix}$ \\ 
$\langle\Sigma_c \bar{D}^*(2S){|\mathcal{E}_{33}|}\Sigma_c \bar{D}^*(2S)\rangle$ & 
$\begin{pmatrix} -2 & 0 \\ 0 & 1 \end{pmatrix}$ & 
$\begin{pmatrix} 1 & 0 & 0 \\ 0 & -2 & 0 \\ 0 & 0 & 1 \end{pmatrix}$ & 
$\begin{pmatrix} -2 & 0 \\ 0 & 1 \end{pmatrix}$ \\ 
$\langle\Sigma_c \bar{D}^*(2S){|\mathcal{F}_{33}|}\Sigma_c \bar{D}^*(2S)\rangle$ & 
$\begin{pmatrix} 0 & -\sqrt{2} \\ -\sqrt{2} & -2 \end{pmatrix}$ & 
$\begin{pmatrix} 0 & 1 & 2 \\ 1 & 0 & -1 \\ 2 & -1 & 0 \end{pmatrix}$ & 
$\begin{pmatrix} 0 & \sqrt{\frac{2}{7}} \\ \sqrt{\frac{2}{7}} & \frac{10}{7} \end{pmatrix}$ \\ \hline
$\langle\Sigma_c \bar{D}^*(2S){|\mathcal{E}_{34}|}\Sigma_c^* \bar{D}^*(2S)\rangle$ & 
$\begin{pmatrix} \sqrt{\frac{2}{3}} & 0 & 0 \\ 0 & \sqrt{\frac{5}{3}} & 0 \end{pmatrix}$ & 
$\begin{pmatrix} \sqrt{\frac{3}{5}} & 0 & 0 & 0 \\  0&\sqrt{\frac{2}{3}} & 0 & 0 \\ 0& 0 & \sqrt{\frac{5}{3}} & 0   \end{pmatrix}$ & 
$\begin{pmatrix} 0 &\sqrt{\frac{2}{3}} & 0 & 0 \\ 0 &0 & \sqrt{\frac{5}{3}} & 0   \end{pmatrix}$ \\ 
$\langle\Sigma_c \bar{D}^*(2S){|\mathcal{F}_{34}|}\Sigma_c^* \bar{D}^*(2S)\rangle$ & 
$\begin{pmatrix} 0 & 4\sqrt{\frac{2}{15}} & \sqrt{\frac{6}{5}} \\ \frac{1}{\sqrt{3}} & - \frac{1}{\sqrt{15}} & \sqrt{\frac{3}{5}}  \end{pmatrix}$ & 
$\begin{pmatrix} 0 &-\frac{1}{\sqrt{6}} & \frac{1}{\sqrt{15}} &\sqrt{\frac{21}{10}} \\ -\frac{4}{\sqrt{15}} & 0 & \frac{4}{\sqrt{15}} & -\sqrt{\frac{6}{35}}\\  \frac{1}{\sqrt{15}} &  \frac{1}{\sqrt{6}} & 0 & \sqrt{\frac{15}{14}} \end{pmatrix}$ & 
$\begin{pmatrix} \sqrt{\frac{2}{5}}  &0 & -4\sqrt{\frac{2}{105}}  & -\frac{4}{\sqrt{35}} \\ -\sqrt{\frac{7}{5}} &-\frac{1}{\sqrt{21}} & \frac{\sqrt{\frac{5}{3}}}{7} & \frac{\sqrt{10}}{7}   \end{pmatrix}$ \\ \hline
$\langle\Sigma_c^* \bar{D}^*(2S){|\mathcal{D}_{44}|}\Sigma_c^* \bar{D}^*(2S)\rangle$ & 
$\begin{pmatrix} 1 & 0 & 0 \\ 0 & 1 & 0 \\ 0 & 0 & 1 \end{pmatrix}$ & 
$\begin{pmatrix} 1 & 0 & 0 & 0 \\ 0 & 1 & 0 & 0 \\ 0 & 0 & 1 & 0 \\ 0 & 0 & 0 & 1 \end{pmatrix}$ & 
$\begin{pmatrix} 1 & 0 & 0 & 0 \\ 0 & 1 & 0 & 0 \\ 0 & 0 & 1 & 0 \\ 0 & 0 & 0 & 1 \end{pmatrix}$ \\ 
$\langle\Sigma_c^* \bar{D}^*(2S){|\mathcal{E}_{44}|}\Sigma_c^* \bar{D}^*(2S)\rangle$ & 
$\begin{pmatrix} \frac{5}{3} & 0 & 0 \\ 0 & \frac{2}{3} & 0 \\ 0 & 0 & -1 \end{pmatrix}$ & 
$\begin{pmatrix} \frac{2}{3} & 0 & 0 & 0 \\ 0 & \frac{5}{3} & 0 & 0 \\ 0 & 0 & \frac{2}{3} & 0 \\ 0 & 0 & 0 & -1 \end{pmatrix}$ & 
$\begin{pmatrix} -1 & 0 & 0 & 0 \\ 0 & \frac{5}{3} & 0 & 0 \\ 0 & 0 & \frac{2}{3} & 0 \\ 0 & 0 & 0 & -1 \end{pmatrix}$ \\ 
$\langle\Sigma_c^* \bar{D}^*(2S){|\mathcal{F}_{44}|}\Sigma_c^* \bar{D}^*(2S)\rangle$ & 
$\begin{pmatrix} 0 & -\frac{7}{3\sqrt{5}} & \frac{2}{\sqrt{5}} \\ -\frac{7}{3\sqrt{5}} & \frac{16}{15} & -\frac{1}{5} \\  \frac{2}{\sqrt{5}} & -\frac{1}{5} &  \frac{8}{5} \end{pmatrix}$ & 
$\begin{pmatrix} 0 &  \frac{7}{3\sqrt{10}} & -\frac{16}{15} & -\frac{\sqrt{\frac{7}{2}}}{5}   \\ \frac{7}{3\sqrt{10}}& 0 & -\frac{7}{3\sqrt{10}} & -\frac{2}{\sqrt{35}} \\ -\frac{16}{15} & -\frac{7}{3\sqrt{10}} & 0 & -\frac{1}{\sqrt{14}} \\ -\frac{\sqrt{\frac{7}{2}}}{5}  & -\frac{2}{\sqrt{35}}& -\frac{1}{\sqrt{14}} & \frac{4}{7} \end{pmatrix}$ & 
$\begin{pmatrix} 0 &\frac{2}{\sqrt{15}}  &\frac{\sqrt{\frac{7}{3}}}{5} & -\frac{2\sqrt{14}}{5} \\ \frac{2}{\sqrt{15}}  & 0 &\frac{\sqrt{\frac{7}{5}}}{3} & -\sqrt{\frac{32}{105}} \\ \frac{\sqrt{\frac{7}{3}}}{5} & \frac{\sqrt{\frac{7}{5}}}{3} & -\frac{16}{21} & -\frac{\sqrt{\frac{2}{3}}}{7} \\ -\frac{2\sqrt{14}}{5} & -\sqrt{\frac{32}{105}} & -\frac{\sqrt{\frac{2}{3}}}{7} & -\frac{4}{7}   \end{pmatrix}$ \\ 
\hline \hline
\end{tabular}%
}
\end{table*}

\begin{table*}[!htbp]
\centering
\caption{The bound-state solutions for the isodoublet coupled $\Sigma_c^{(*)}\bar{D}^{(*)}(2S)$ molecules, including the binding energy \(E\), root-mean-square radius \(r_{\text{RMS}}\), and the channel probabilities \(p_i\). The cutoff parameter $\Lambda$, $E$, and $r_{\rm RMS}$ are given in units of GeV, MeV, and fm, respectively.}
\label{tab:result}
\renewcommand\tabcolsep{0.15cm}
\renewcommand{\arraystretch}{2.3}

\begin{tabular}{c | c c c c c c}
\hline\hline

$I(J^P)=1/2(1/2^{-})$ &
$\Lambda$ & $E$ & $r_{\rm RMS}$ & $\Sigma_c\bar{D}(2S)\left[^2S_\frac{1}{2} \right]$ & $\Sigma_c\bar{D}^*(2S)\left[^2S_\frac{1}{2}/^4D_\frac{1}{2} \right]$ & $\Sigma_c^*\bar{D}^*(2S)\left[^2S_\frac{1}{2}/^4D_\frac{1}{2}/^6D_\frac{1}{2} \right]$\\
\hline

\multirow{3}{*}{$\Sigma_c\bar D(2S)/\Sigma_c\bar D^*(2S)/\Sigma_c^*\bar D^*(2S)$}
&0.90&$-5.02$&1.48&
\textbf{73.84}&
\textbf{24.66}/0.51&
\textbf{0.60}/0.02/0.37\\

&0.93&$-9.73$&1.02&
\textbf{63.54}&
\textbf{35.16}/0.42&
\textbf{0.58}/0.02/0.70\\

&0.95&$-14.61$&0.85&
\textbf{55.68}&
\textbf{43.30}/0.33&
\textbf{0.35}/0.03/0.31\\

\cline{2-7}

\multirow{3}{*}{$\Sigma_c\bar D^*(2S)/\Sigma_c^*\bar D^*(2S)$}
&0.84&$-5.08$&1.49&
\textbf{/}&
\textbf{94.79}/0.77&
\textbf{4.09}/0.17/0.18\\

&0.87&$-9.90$&1.16&
\textbf{/}&
\textbf{91.54}/0.75&
\textbf{7.32}/0.16/0.16\\

&0.89&$-14.58$&0.95&
\textbf{/}&
\textbf{88.78}/0.72&
\textbf{10.10}/0.14/0.19\\

\hline

$I(J^P)=1/2(3/2^{-})$ &
$\Lambda$ & $E$ & $r_{\rm RMS}$ &$\Sigma_c^*\bar{D}(2S)\left[{}^4{S}_{\frac{3}{2}}/{}^2{D}_{\frac{3}{2}} \right]$ & $\Sigma_c\bar{D}^*(2S)\left[{}^4{S}_{\frac{3}{2}}/{}^2{D}_{\frac{3}{2}}/ {}^4{D}_{\frac{3}{2}} \right]$ &  $\Sigma_c^*\bar{D}^*(2S)\left[{}^4{S}_{\frac{3}{2}}/{}^2{D}_{\frac{3}{2}}/ {}^4{D}_{\frac{3}{2}}/ {}^6{D}_{\frac{3}{2}} \right]$\\
\hline

\multirow{3}{*}{$\Sigma_c\bar D^*(2S)/\Sigma_c^*\bar D(2S)/\Sigma_c^*\bar D^*(2S)$}
&1.21&$-4.87$&1.63&
\textbf{12.33}/1.11&
\textbf{78.11}/0.21/4.57&
\textbf{1.85}/0.01/0.13/1.66\\

&1.29&$-10.13$&1.22&
\textbf{17.98}/1.31&
\textbf{70.62}/0.15/4.94&
\textbf{2.52}/0.02/0.25/2.21\\

&1.34&$-14.27$&1.08&
\textbf{20.32}/1.36&
\textbf{67.63}/0.12/5.01&
\textbf{2.78}/0.03/0.30/2.45\\

\cline{2-7}

\multirow{3}{*}{$\Sigma_c^*\bar D(2S)/\Sigma_c^*\bar D^*(2S)$}
&0.92&$-5.70$&1.43&
\textbf{77.02}/0.15&
\textbf{/}&
\textbf{21.72}/0.01/0.12/0.98\\

&0.94&$-10.24$&1.02&
\textbf{70.29}/0.19&
\textbf{/}&
\textbf{28.41}/0.02/0.08/1.01\\

&0.96&$-14.97$&0.83&

\textbf{65.94}/0.22&
\textbf{/}&
\textbf{32.78}/0.02/0.06/0.98\\

\hline\hline
\end{tabular}
\end{table*}

\end{document}